

%
%
%
\def\unredoffs{} \def\redoffs{\voffset=-.31truein\hoffset=-.59truein}
\def\speclscape{\special{ps: landscape}}
%
%
%
%
\newbox\leftpage \newdimen\fullhsize \newdimen\hstitle \newdimen\hsbody
\tolerance=1000\hfuzz=2pt
\catcode`\@=11 
\def\bigans{b }
\def\answ{b }

%
\ifx\answ\bigans\message{(This will come out unreduced.}
\magnification=1200\unredoffs\baselineskip=16pt plus 2pt minus 1pt
\hsbody=\hsize \hstitle=\hsize 
\else\message{(This will be reduced.} \let\l@r=L
\magnification=1000\baselineskip=16pt plus 2pt minus 1pt \vsize=7truein
\redoffs \hstitle=8truein\hsbody=4.75truein\fullhsize=10truein\hsize=\hsbody
\output={\ifnum\pageno=0 
  \shipout\vbox{\speclscape{\hsize\fullhsize\makeheadline}
    \hbox to \fullhsize{\hfill\pagebody\hfill}}\advancepageno
  \else
  \almostshipout{\leftline{\vbox{\pagebody\makefootline}}}\advancepageno
  \fi}
\def\almostshipout#1{\if L\l@r \count1=1 \message{[\the\count0.\the\count1]}
      \global\setbox\leftpage=#1 \global\let\l@r=R
 \else \count1=2
  \shipout\vbox{\speclscape{\hsize\fullhsize\makeheadline}
      \hbox to\fullhsize{\box\leftpage\hfil#1}}  \global\let\l@r=L\fi}
\fi
%
\newcount\yearltd\yearltd=\year\advance\yearltd by -1900

\def\Title#1#2{\nopagenumbers\abstractfont\hsize=\hstitle\rightline{#1}%
\vskip 1in\centerline{\titlefont #2}\abstractfont\vskip .5in\pageno=0}
\def\Date#1{\vfill\leftline{#1}\tenpoint\supereject\global\hsize=\hsbody%
\footline={\hss\tenrm\folio\hss}}
%

\def\draftmode{\message{ DRAFTMODE }\def\draftdate{{\rm preliminary draft:
\number\month/\number\day/\number\yearltd\ \ \hourmin}}%
\headline={\hfil\draftdate}\writelabels\baselineskip=20pt plus 2pt minus 2pt
 {\count255=\time\divide\count255 by 60 \xdef\hourmin{\number\count255}
  \multiply\count255 by-60\advance\count255 by\time
  \xdef\hourmin{\hourmin:\ifnum\count255<10 0\fi\the\count255}}}
\def\nolabels{\def\wrlabeL##1{}\def\eqlabeL##1{}\def\reflabeL##1{}}
\def\writelabels{\def\wrlabeL##1{\leavevmode\vadjust{\rlap{\smash%
{\line{{\escapechar=` \hfill\rlap{\sevenrm\hskip.03in\string##1}}}}}}}%
\def\eqlabeL##1{{\escapechar-1\rlap{\sevenrm\hskip.05in\string##1}}}%
\def\reflabeL##1{\noexpand\llap{\noexpand\sevenrm\string\string\string##1}}}
\nolabels
%
\global\newcount\secno \global\secno=0
\global\newcount\meqno \global\meqno=1
\def\newsec#1{\global\advance\secno by1\message{(\the\secno. #1)}
\global\subsecno=0\eqnres@t\noindent{\bf\the\secno. #1}
\writetoca{{\secsym} {#1}}\par\nobreak\medskip\nobreak}
\def\eqnres@t{\xdef\secsym{\the\secno.}\global\meqno=1\bigbreak\bigskip}
\def\sequentialequations{\def\eqnres@t{\bigbreak}}\xdef\secsym{}
\global\newcount\subsecno \global\subsecno=0
\def\subsec#1{\global\advance\subsecno by1\message{(\secsym\the\subsecno. #1)}
\ifnum\lastpenalty>9000\else\bigbreak\fi
\noindent{\it\secsym\the\subsecno. #1}\writetoca{\string\quad
{\secsym\the\subsecno.} {#1}}\par\nobreak\medskip\nobreak}
\def\appendix#1#2{\global\meqno=1\global\subsecno=0\xdef\secsym{\hbox{#1.}}
\bigbreak\bigskip\noindent{\bf Appendix #1. #2}\message{(#1. #2)}
\writetoca{Appendix {#1.} {#2}}\par\nobreak\medskip\nobreak}
%
%
\def\eqnn#1{\xdef #1{(\secsym\the\meqno)}\writedef{#1\leftbracket#1}%
\global\advance\meqno by1\wrlabeL#1}
\def\eqna#1{\xdef #1##1{\hbox{$(\secsym\the\meqno##1)$}}
\writedef{#1\numbersign1\leftbracket#1{\numbersign1}}%
\global\advance\meqno by1\wrlabeL{#1$\{\}$}}
\def\eqn#1#2{\xdef #1{(\secsym\the\meqno)}\writedef{#1\leftbracket#1}%
\global\advance\meqno by1$$#2\eqno#1\eqlabeL#1$$}
%
\newskip\footskip\footskip14pt plus 1pt minus 1pt 
\def\footnotefont{\ninepoint}\def\f@t#1{\footnotefont #1\@foot}
\def\f@@t{\baselineskip\footskip\bgroup\footnotefont\aftergroup\@foot\let\next}
\setbox\strutbox=\hbox{\vrule height9.5pt depth4.5pt width0pt}
\global\newcount\ftno \global\ftno=0
\def\foot{\global\advance\ftno by1\footnote{$^{\the\ftno}$}}
%
\newwrite\ftfile
\def\footend{\def\foot{\global\advance\ftno by1\chardef\wfile=\ftfile
$^{\the\ftno}$\ifnum\ftno=1\immediate\openout\ftfile=foots.tmp\fi%
\immediate\write\ftfile{\noexpand\smallskip%
\noexpand\item{f\the\ftno:\ }\pctsign}\findarg}%
\def\footatend{\vfill\eject\immediate\closeout\ftfile{\parindent=20pt
\centerline{\bf Footnotes}\nobreak\bigskip\input foots.tmp }}}
\def\footatend{}
%
%
\global\newcount\refno \global\refno=1
\newwrite\rfile
\def\ref{[\the\refno]\nref}
\def\nref#1{\xdef#1{[\the\refno]}\writedef{#1\leftbracket#1}%
\ifnum\refno=1\immediate\openout\rfile=refs.tmp\fi
\global\advance\refno by1\chardef\wfile=\rfile\immediate
\write\rfile{\noexpand\item{#1\ }\reflabeL{#1\hskip.31in}\pctsign}\findarg}
\def\findarg#1#{\begingroup\obeylines\newlinechar=`\^^M\pass@rg}
{\obeylines\gdef\pass@rg#1{\writ@line\relax #1^^M\hbox{}^^M}%
\gdef\writ@line#1^^M{\expandafter\toks0\expandafter{\striprel@x #1}%
\edef\next{\the\toks0}\ifx\next\em@rk\let\next=\endgroup\else\ifx\next\empty%
\else\immediate\write\wfile{\the\toks0}\fi\let\next=\writ@line\fi\next\relax}}
\def\striprel@x#1{} \def\em@rk{\hbox{}}
\def\lref{\begingroup\obeylines\lr@f}
\def\lr@f#1#2{\gdef#1{\ref#1{#2}}\endgroup\unskip}

\def\addref#1{\immediate\write\rfile{\noexpand\item{}#1}} 
\def\startrefs#1{\immediate\openout\rfile=refs.tmp\refno=#1}
\def\xref{\expandafter\xr@f}\def\xr@f[#1]{#1}
\def\refs#1{\count255=1[\r@fs #1{\hbox{}}]}
\def\r@fs#1{\ifx\und@fined#1\message{reflabel \string#1 is undefined.}%
\nref#1{need to supply reference \string#1.}\fi%
\vphantom{\hphantom{#1}}\edef\next{#1}\ifx\next\em@rk\def\next{}%
\else\ifx\next#1\ifodd\count255\relax\xref#1\count255=0\fi%
\else#1\count255=1\fi\let\next=\r@fs\fi\next}
%

%
\newwrite\ffile\global\newcount\figno \global\figno=1
\def\fig{fig.~\the\figno\nfig}
\def\nfig#1{\xdef#1{fig.~\the\figno}%
\writedef{#1\leftbracket fig.\noexpand~\the\figno}%
\ifnum\figno=1\immediate\openout\ffile=figs.tmp\fi\chardef\wfile=\ffile%
\immediate\write\ffile{\noexpand\medskip\noexpand\item{Fig.\ \the\figno. }
\reflabeL{#1\hskip.55in}\pctsign}\global\advance\figno by1\findarg}
\def\xfig{\expandafter\xf@g}\def\xf@g fig.\penalty\@M\ {}
\def\figs#1{figs.~\f@gs #1{\hbox{}}}
\def\f@gs#1{\edef\next{#1}\ifx\next\em@rk\def\next{}\else
\ifx\next#1\xfig #1\else#1\fi\let\next=\f@gs\fi\next}
\newwrite\lfile
{\escapechar-1\xdef\pctsign{\string\%}\xdef\leftbracket{\string\{}
\xdef\rightbracket{\string\}}\xdef\numbersign{\string\#}}

\def\writestop{\def\writestoppt{\immediate\write\lfile{\string\pageno%
\the\pageno\string\startrefs\leftbracket\the\refno\rightbracket%
\string\def\string\secsym\leftbracket\secsym\rightbracket%
\string\secno\the\secno\string\meqno\the\meqno}\immediate\closeout\lfile}}
\def\writestoppt{}\def\writedef#1{}
\def\seclab#1{\xdef #1{\the\secno}\writedef{#1\leftbracket#1}\wrlabeL{#1=#1}}
\def\subseclab#1{\xdef #1{\secsym\the\subsecno}%
\writedef{#1\leftbracket#1}\wrlabeL{#1=#1}}
\newwrite\tfile \def\writetoca#1{}
\def\leaderfill{\leaders\hbox to 1em{\hss.\hss}\hfill}
\def\writetoc{\immediate\openout\tfile=toc.tmp
   \def\writetoca##1{{\edef\next{\write\tfile{\noindent ##1
   \string\leaderfill {\noexpand\number\pageno} \par}}\next}}}

%
\catcode`\@=12 
%
\edef\tfontsize{\ifx\answ\bigans scaled\magstep3\else scaled\magstep4\fi}
\font\titlerm=cmr10 \tfontsize \font\titlerms=cmr7 \tfontsize
\font\titlermss=cmr5 \tfontsize \font\titlei=cmmi10 \tfontsize
\font\titleis=cmmi7 \tfontsize \font\titleiss=cmmi5 \tfontsize
\font\titlesy=cmsy10 \tfontsize \font\titlesys=cmsy7 \tfontsize
\font\titlesyss=cmsy5 \tfontsize \font\titleit=cmti10 \tfontsize
\skewchar\titlei='177 \skewchar\titleis='177 \skewchar\titleiss='177
\skewchar\titlesy='60 \skewchar\titlesys='60 \skewchar\titlesyss='60
\def\titlefont{\def\rm{\fam0\titlerm}
\textfont0=\titlerm \scriptfont0=\titlerms \scriptscriptfont0=\titlermss
\textfont1=\titlei \scriptfont1=\titleis \scriptscriptfont1=\titleiss
\textfont2=\titlesy \scriptfont2=\titlesys \scriptscriptfont2=\titlesyss
\textfont\itfam=\titleit \def\it{\fam\itfam\titleit}\rm}
 \ifx\answ\bigans\else scaled\magstep1\fi
\ifx\answ\bigans\def\abstractfont{\tenpoint}\else
\font\abssl=cmsl10 scaled \magstep1
\font\absrm=cmr10 scaled\magstep1 \font\absrms=cmr7 scaled\magstep1
\font\absrmss=cmr5 scaled\magstep1 \font\absi=cmmi10 scaled\magstep1
\font\absis=cmmi7 scaled\magstep1 \font\absiss=cmmi5 scaled\magstep1
\font\abssy=cmsy10 scaled\magstep1 \font\abssys=cmsy7 scaled\magstep1
\font\abssyss=cmsy5 scaled\magstep1 \font\absbf=cmbx10 scaled\magstep1
\skewchar\absi='177 \skewchar\absis='177 \skewchar\absiss='177
\skewchar\abssy='60 \skewchar\abssys='60 \skewchar\abssyss='60
\def\abstractfont{\def\rm{\fam0\absrm}
\textfont0=\absrm \scriptfont0=\absrms \scriptscriptfont0=\absrmss
\textfont1=\absi \scriptfont1=\absis \scriptscriptfont1=\absiss
\textfont2=\abssy \scriptfont2=\abssys \scriptscriptfont2=\abssyss
\textfont\itfam=\bigit \def\it{\fam\itfam\bigit}\def\footnotefont{\tenpoint}%
\textfont\slfam=\abssl \def\sl{\fam\slfam\abssl}%
\textfont\bffam=\absbf \def\bf{\fam\bffam\absbf}\rm}\fi
\def\tenpoint{\def\rm{\fam0\tenrm}
\textfont0=\tenrm \scriptfont0=\sevenrm \scriptscriptfont0=\fiverm
\textfont1=\teni  \scriptfont1=\seveni  \scriptscriptfont1=\fivei
\textfont2=\tensy \scriptfont2=\sevensy \scriptscriptfont2=\fivesy
\textfont\itfam=\tenit \def\it{\fam\itfam\tenit}\def\footnotefont{\ninepoint}%
\textfont\bffam=\tenbf \def\bf{\fam\bffam\tenbf}\def\sl{\fam\slfam\tensl}\rm}
\font\ninerm=cmr9 \font\sixrm=cmr6 \font\ninei=cmmi9 \font\sixi=cmmi6
\font\ninesy=cmsy9 \font\sixsy=cmsy6 \font\ninebf=cmbx9
\font\nineit=cmti9 \font\ninesl=cmsl9 \skewchar\ninei='177
\skewchar\sixi='177 \skewchar\ninesy='60 \skewchar\sixsy='60
\def\ninepoint{\def\rm{\fam0\ninerm}
\textfont0=\ninerm \scriptfont0=\sixrm \scriptscriptfont0=\fiverm
\textfont1=\ninei \scriptfont1=\sixi \scriptscriptfont1=\fivei
\textfont2=\ninesy \scriptfont2=\sixsy \scriptscriptfont2=\fivesy
\textfont\itfam=\ninei \def\it{\fam\itfam\nineit}\def\sl{\fam\slfam\ninesl}%
\textfont\bffam=\ninebf \def\bf{\fam\bffam\ninebf}\rm}
%
%

\hyphenation{anom-aly anom-alies coun-ter-term coun-ter-terms}
\def\inv{^{\raise.15ex\hbox{${\scriptscriptstyle -}$}\kern-.05em 1}}

\def\Dsl{\,\raise.15ex\hbox{/}\mkern-13.5mu D} 
\def\dsl{\raise.15ex\hbox{/}\kern-.57em\partial}

\def\tr{{\rm tr}} 
\font\bigit=cmti10 scaled \magstep1
\def\lspace{\ifx\answ\bigans{}\else\qquad\fi}
\def\lbspace{\ifx\answ\bigans{}\else\hskip-.2in\fi} 
\def\boxeqn#1{\vcenter{\vbox{\hrule\hbox{\vrule\kern3pt\vbox{\kern3pt
    \hbox{${\displaystyle #1}$}\kern3pt}\kern3pt\vrule}\hrule}}}
\def\mbox#1#2{\vcenter{\hrule \hbox{\vrule height#2in
        \kern#1in \vrule} \hrule}}  
%

\def\darr#1{\raise1.5ex\hbox{$\leftrightarrow$}\mkern-16.5mu #1}

\def\roughly#1{\raise.3ex\hbox{$#1$\kern-.75em\lower1ex\hbox{$\sim$}}}

\let\includefigures=\iftrue
\let\useblackboard=\iftrue
\newfam\black

\includefigures
\message{If you do not have epsf.tex (to include figures),}
\message{change the option at the top of the tex file.}
\input epsf
\def\figin{\epsfcheck\figin}\def\figins{\epsfcheck\figins}
\def\epsfcheck{\ifx\epsfbox\UnDeFiNeD
\message{(NO epsf.tex, FIGURES WILL BE IGNORED)}
\gdef\figin##1{\vskip2in}\gdef\figins##1{\hskip.5in}
\else\message{(FIGURES WILL BE INCLUDED)}%
\gdef\figin##1{##1}\gdef\figins##1{##1}\fi}
\def\DefWarn#1{}
\def\figinsert{\goodbreak\midinsert}
\def\ifig#1#2#3{\DefWarn#1\xdef#1{fig.~\the\figno}
\writedef{#1\leftbracket fig.\noexpand~\the\figno}%
\figinsert\figin{\centerline{#3}}\medskip\centerline{\vbox{
\baselineskip12pt\advance\hsize by -1truein
\noindent\footnotefont{\bf Fig.~\the\figno:} #2}}
\endinsert\global\advance\figno by1}
\else
\def\ifig#1#2#3{\xdef#1{fig.~\the\figno}
\writedef{#1\leftbracket fig.\noexpand~\the\figno}%
\global\advance\figno by1} \fi

\def\id{{1 \kern-.28em {\rm l}}}

\def\K3{{\bf K3}}
\def\journal#1&#2(#3){\unskip, \sl #1\ \bf #2 \rm(19#3) }
\def\andjournal#1&#2(#3){\sl #1~\bf #2 \rm (19#3) }

\def\hat{\widehat}
\def\ie{{\it i.e.}}

\def\tilde{\widetilde}

\def\frac#1#2{{#1\over#2}}

\def\inbar{\,\vrule height1.5ex width.4pt depth0pt}
\def\IC{\relax\hbox{$\inbar\kern-.3em{\rm C}$}}
\def\IR{\relax{\rm I\kern-.18em R}}
\def\IP{\relax{\rm I\kern-.18em P}}

%
%

%
\catcode`\@=11
\def\slash#1{\mathord{\mathpalette\c@ncel{#1}}}
\overfullrule=0pt

\def\DD{{\cal D}}

\def\LL{{\cal L}}

\def\OO{{\cal O}}

\def\RR{{\cal R}}
\def\SS{{\cal S}}

\def\underrel#1\over#2{\mathrel{\mathop{\kern\z@#1}\limits_{#2}}}

\catcode`\@=12


%

\def\tr{{\rm tr}}


\def\p{{\partial}}

\def\tc{{\tilde c}}

\def\LL{{\cal L}}


\lref\MetsaevTseytlina{
  R.~R.~Metsaev and A.~A.~Tseytlin,
  ``Curvature Cubed Terms in String Theory Effective Actions,''
  Phys.\ Lett.\  B {\bf 185}, 52 (1987).
}

\lref\BastianelliFTa{
  F.~Bastianelli, S.~Frolov and A.~A.~Tseytlin,
  ``Three-point correlators of stress tensors in maximally-supersymmetric
  conformal theories in d = 3 and d = 6,''
  Nucl.\ Phys.\  B {\bf 578}, 139 (2000)
  [arXiv:hep-th/9911135].
}

\lref\BastianelliFTb{
  F.~Bastianelli, S.~Frolov and A.~A.~Tseytlin,
  ``Conformal anomaly of (2,0) tensor multiplet in six dimensions and  AdS/CFT
  correspondence,''
  JHEP {\bf 0002}, 013 (2000)
  [arXiv:hep-th/0001041].
}


\lref\BuchelMyers{
  A.~Buchel and R.~C.~Myers,
  ``Causality of Holographic Hydrodynamics,''
  arXiv:0906.2922 [hep-th].
}

\lref\BriganteLMSYa{
  M.~Brigante, H.~Liu, R.~C.~Myers, S.~Shenker and S.~Yaida,
  ``Viscosity Bound Violation in Higher Derivative Gravity,''
  Phys.\ Rev.\  D {\bf 77}, 126006 (2008)
  [arXiv:0712.0805 [hep-th]].
}

\lref\BriganteLMSYb{
  M.~Brigante, H.~Liu, R.~C.~Myers, S.~Shenker and S.~Yaida,
  ``The Viscosity Bound and Causality Violation,''
  Phys.\ Rev.\ Lett.\  {\bf 100}, 191601 (2008)
  [arXiv:0802.3318 [hep-th]].
}


\lref\Hofman{
  D.~M.~Hofman,
  ``Higher Derivative Gravity, Causality and Positivity of Energy in a UV
  complete QFT,''
  arXiv:0907.1625 [hep-th].
}

\lref\HofmanMaldacena{
  D.~M.~Hofman and J.~Maldacena,
  ``Conformal collider physics: Energy and charge correlations,''
  JHEP {\bf 0805}, 012 (2008)
  [arXiv:0803.1467 [hep-th]].
}


\lref\DehghaniPourhasan{
  M.~H.~Dehghani and R.~Pourhasan,
  ``Thermodynamic instability of black holes of third order Lovelock gravity,''
  Phys.\ Rev.\  D {\bf 79}, 064015 (2009)
  [arXiv:0903.4260 [gr-qc]].
}

\lref\DehghaniBH{
  M.~H.~Dehghani, N.~Bostani and S.~H.~Hendi,
  ``Magnetic Branes in Third Order Lovelock-Born-Infeld Gravity,''
  Phys.\ Rev.\  D {\bf 78}, 064031 (2008)
  [arXiv:0806.1429 [gr-qc]].
}

\lref\DehghaniAH{
  M.~H.~Dehghani, N.~Alinejadi and S.~H.~Hendi,
  ``Topological Black Holes in Lovelock-Born-Infeld Gravity,''
  Phys.\ Rev.\  D {\bf 77}, 104025 (2008)
  [arXiv:0802.2637 [hep-th]].
}

\lref\Lovelock{
  D.~Lovelock
  ``The Einstein Tensor and Its Generalizations''
  J.Math.Phys. \ {\bf 12}, 498 (1971)
}

\lref\NojiriMH{
  S.~Nojiri and S.~D.~Odintsov,
  ``On the conformal anomaly from higher derivative gravity in AdS/CFT
  correspondence,''
  Int.\ J.\ Mod.\ Phys.\  A {\bf 15}, 413 (2000)
  [arXiv:hep-th/9903033].
}

\lref\HS{
  M.~Henningson and K.~Skenderis,
  ``The holographic Weyl anomaly,''
  JHEP {\bf 9807}, 023 (1998)
  [arXiv:hep-th/9806087];
  ``Holography and the Weyl anomaly,''
  Fortsch.\ Phys.\  {\bf 48}, 125 (2000)
  [arXiv:hep-th/9812032].
}

\lref\OsbornErdmenger{
  J.~Erdmenger and H.~Osborn,
  ``Conserved currents and the energy-momentum tensor in conformally  invariant
  theories for general dimensions,''
  Nucl.\ Phys.\  B {\bf 483}, 431 (1997)
  [arXiv:hep-th/9605009].
}

\lref\OsbornPetkou{
  H.~Osborn and A.~C.~Petkou,
  ``Implications of Conformal Invariance in Field Theories for General
  Dimensions,''
  Annals Phys.\  {\bf 231}, 311 (1994)
  [arXiv:hep-th/9307010].
}

\lref\BanerjeeFM{
  N.~Banerjee and S.~Dutta,
  ``Shear Viscosity to Entropy Density Ratio in Six Derivative Gravity,''
  JHEP {\bf 0907}, 024 (2009)
  [arXiv:0903.3925 [hep-th]].
}

\lref\BuchelVZ{
  A.~Buchel, R.~C.~Myers and A.~Sinha,
  ``Beyond eta/s = 1/4pi,''
  JHEP {\bf 0903}, 084 (2009)
  [arXiv:0812.2521 [hep-th]].
}

\lref\AdamsZK{
  A.~Adams, A.~Maloney, A.~Sinha and S.~E.~Vazquez,
  ``1/N Effects in Non-Relativistic Gauge-Gravity Duality,''
  JHEP {\bf 0903}, 097 (2009)
  [arXiv:0812.0166 [hep-th]].
}

\lref\ShuAX{
  F.~W.~Shu,
  ``The Quantum Viscosity Bound In Lovelock Gravity,''
  arXiv:0910.0607 [hep-th].
}

\lref\GeAC{
  X.~H.~Ge, S.~J.~Sin, S.~F.~Wu and G.~H.~Yang,
  ``Shear viscosity and instability from third order Lovelock gravity,''
  arXiv:0905.2675 [hep-th].
}

\lref\BanerjeeFM{
  N.~Banerjee and S.~Dutta,
  ``Shear Viscosity to Entropy Density Ratio in Six Derivative Gravity,''
  JHEP {\bf 0907}, 024 (2009)
  [arXiv:0903.3925 [hep-th]].
}

\lref\CremoniniSY{
  S.~Cremonini, K.~Hanaki, J.~T.~Liu and P.~Szepietowski,
  ``Higher derivative effects on eta/s at finite chemical potential,''
  Phys.\ Rev.\  D {\bf 80}, 025002 (2009)
  [arXiv:0903.3244 [hep-th]].
}

\lref\GeEH{
  X.~H.~Ge and S.~J.~Sin,
  ``Shear viscosity, instability and the upper bound of the Gauss-Bonnet
  coupling constant,''
  JHEP {\bf 0905}, 051 (2009)
  [arXiv:0903.2527 [hep-th]].
}

\lref\BrusteinCG{
  R.~Brustein and A.~J.~M.~Medved,
  ``The ratio of shear viscosity to entropy density in generalized theories of
  gravity,''
  Phys.\ Rev.\  D {\bf 79}, 021901 (2009)
  [arXiv:0808.3498 [hep-th]].
}

\lref\KatsMQ{
  Y.~Kats and P.~Petrov,
  ``Effect of curvature squared corrections in AdS on the viscosity of the dual
  gauge theory,''
  JHEP {\bf 0901}, 044 (2009)
  [arXiv:0712.0743 [hep-th]].
}

\lref\KSS{
  P.~Kovtun, D.~T.~Son and A.~O.~Starinets,
  ``Viscosity in strongly interacting quantum field theories from black hole
  physics,''
  Phys.\ Rev.\ Lett.\  {\bf 94}, 111601 (2005)
  [arXiv:hep-th/0405231].
  ``Holography and hydrodynamics: Diffusion on stretched horizons,''
  JHEP {\bf 0310}, 064 (2003)
  [arXiv:hep-th/0309213].
}

\lref\NeupaneDC{
  I.~P.~Neupane and N.~Dadhich,
  ``Higher Curvature Gravity: Entropy Bound and Causality Violation,''
  arXiv:0808.1919 [hep-th].
}

\lref\BuchelVZ{
  A.~Buchel, R.~C.~Myers and A.~Sinha,
  ``Beyond eta/s = 1/4pi,''
  JHEP {\bf 0903}, 084 (2009)
  [arXiv:0812.2521 [hep-th]].
}

\lref\PalQG{
  S.~S.~Pal,
  ``$\eta/s$ at finite coupling,''
  arXiv:0910.0101 [hep-th].
}

\lref\WP{
Work in progress.  }

\lref\SinhaMyers{
   A.~Sinha and R.~Myers
  ``The viscosity bound in string theory,''
   arXiv:0907.4798 [hep-th]
}

\lref\SveshikovTkachov{
  N.~A.~Sveshnikov and F.~V.~Tkachov,
  ``Jets and quantum field theory,''
  Phys.\ Lett.\  B {\bf 382} (1996) 403
  [arXiv:hep-ph/9512370].
}

\lref\ShapereTachikawa{
  A.~D.~Shapere and Y.~Tachikawa,
  ``Central charges of N=2 superconformal field theories in four dimensions,''
  JHEP {\bf 0809}, 109 (2008)
  [arXiv:0804.1957 [hep-th]].
}

\lref\BoulwareDeser{
  D.~G.~Boulware and S.~Deser,
  ``String Generated Gravity Models,''
  Phys.\ Rev.\ Lett.\  {\bf 55}, 2656 (1985).
}

\lref\Myers{
  R.~C.~Myers,
  ``HIGHER DERIVATIVE GRAVITY, SURFACE TERMS AND STRING THEORY,''
  Phys.\ Rev.\  D {\bf 36}, 392 (1987).
}

\lref\Zumino{
  B.~Zumino,
  ``Gravity Theories In More Than Four-Dimensions,''
  Phys.\ Rept.\  {\bf 137}, 109 (1986).
}

\lref\Cai{
  R.~G.~Cai,
  ``Gauss-Bonnet black holes in AdS spaces,''
  Phys.\ Rev.\  D {\bf 65}, 084014 (2002)
  [arXiv:hep-th/0109133].
}

\lref\ExirifardJabbari{
  Q.~Exirifard and M.~M.~Sheikh-Jabbari,
  ``Lovelock Gravity at the Crossroads of Palatini and Metric Formulations,''
  Phys.\ Lett.\  B {\bf 661}, 158 (2008)
  [arXiv:0705.1879 [hep-th]].
}

\lref\Cremoninia{
  S.~Cremonini, J.~T.~Liu and P.~Szepietowski,
  ``Higher Derivative Corrections to R-charged Black Holes: Boundary
  Counterterms and the Mass-Charge Relation,''
  arXiv:0910.5159 [hep-th].
}

\lref\Paulos{
  M.~F.~Paulos,
  ``Transport coefficients, membrane couplings and universality at
 extremality,''
  arXiv:0910.4602 [hep-th].
}

\lref\Caib{
  R.~G.~Cai, Y.~Liu and Y.~W.~Sun,
  ``Transport Coefficients from Extremal Gauss-Bonnet Black Holes,''
  arXiv:0910.4705 [hep-th].
}

\lref\Ohta{R.~G.~Cai, Z.~Y.~Nie, N.~Ohta and Y.~W.~Sun,
  ``Shear Viscosity from Gauss-Bonnet Gravity with a Dilaton Coupling,''
  Phys.\ Rev.\  D {\bf 79} (2009) 066004
  [arXiv:0901.1421 [hep-th]].
}

\lref\Solodukhina{S.~N.~Solodukhin,
  ``Entanglement entropy, conformal invariance and extrinsic geometry,''
  Phys.\ Lett.\  B {\bf 665}, 305 (2008)
  [arXiv:0802.3117 [hep-th]].
}

\lref\Fursaeva{D.~V.~Fursaev,
  ``Proof of the holographic formula for entanglement entropy,''
  JHEP {\bf 0609}, 018 (2006)
  [arXiv:hep-th/0606184].
}

\lref\Takayanagia{S.~Ryu and T.~Takayanagi,
  ``Aspects of holographic entanglement entropy,''
  JHEP {\bf 0608}, 045 (2006)
  [arXiv:hep-th/0605073].
}

\lref\FursaevSolodukhina{D.~V.~Fursaev and S.~N.~Solodukhin,
  ``On The Description Of The Riemannian Geometry In The Presence Of Conical
  Defects,''
  Phys.\ Rev.\  D {\bf 52}, 2133 (1995)
  [arXiv:hep-th/9501127].
}

\lref\CasiniHuertaa{H.~Casini and M.~Huerta,
  ``Universal terms for the entanglement entropy in 2+1 dimensions,''
  Nucl.\ Phys.\  B {\bf 764}, 183 (2007)
  [arXiv:hep-th/0606256].
}

\lref\CasiniHuertaReview{H.~Casini and M.~Huerta,
  ``Entanglement entropy in free quantum field theory,''
  J.\ Phys.\ A  {\bf 42}, 504007 (2009)
  [arXiv:0905.2562 [hep-th]].
}

\lref\Bucheletala{
  A.~Buchel, J.~Escobedo, R.~C.~Myers, M.~F.~Paulos, A.~Sinha and M.~Smolkin,
  ``Holographic GB gravity in arbitrary dimensions,''
  JHEP {\bf 1003}, 111 (2010)
  [arXiv:0911.4257 [hep-th]].
}

\lref\Bianchietal{
  M.~Bianchi, D.~Z.~Freedman and K.~Skenderis,
  ``How to go with an RG flow,''
  JHEP {\bf 0108}, 041 (2001)
  [arXiv:hep-th/0105276].
}

\lref\Klebanovetal{
  I.~R.~Klebanov, D.~Kutasov and A.~Murugan,
  ``Entanglement as a Probe of Confinement,''
  Nucl.\ Phys.\  B {\bf 796}, 274 (2008)
  [arXiv:0709.2140 [hep-th]].
}

\lref\Girardelloa{
  L.~Girardello, M.~Petrini, M.~Porrati and A.~Zaffaroni,
  ``Novel local CFT and exact results on perturbations of N = 4 super
  Yang-Mills from AdS dynamics,''
  JHEP {\bf 9812}, 022 (1998)
  [arXiv:hep-th/9810126].
}

\lref\Girardellob{
  L.~Girardello, M.~Petrini, M.~Porrati and A.~Zaffaroni,
  ``The supergravity dual of N = 1 super Yang-Mills theory,''
  Nucl.\ Phys.\  B {\bf 569}, 451 (2000)
  [arXiv:hep-th/9909047].
}

\lref\Sinha{
  A.~Sinha,
  ``On the new massive gravity and AdS/CFT,''
  arXiv:1003.0683 [hep-th].
}

\lref\CamanhoEdelsteina{
  X.~O.~Camanho and J.~D.~Edelstein,
  ``Causality constraints in AdS/CFT from conformal collider physics and
  Gauss-Bonnet gravity,''
  JHEP {\bf 1004}, 007 (2010)
  [arXiv:0911.3160 [hep-th]].
}

\lref\DKPa{
  J.~de Boer, M.~Kulaxizi and A.~Parnachev,
  ``AdS(7)/CFT(6), Gauss-Bonnet Gravity, and Viscosity Bound,''
  JHEP {\bf 1003}, 087 (2010)
  [arXiv:0910.5347 [hep-th]].
}

\lref\Takayanagib{
  T.~Hirata and T.~Takayanagi,
  ``AdS/CFT and strong subadditivity of entanglement entropy,''
  JHEP {\bf 0702}, 042 (2007)
  [arXiv:hep-th/0608213].
}

 \lref\Lohmayera{
  R.~Lohmayer, H.~Neuberger, A.~Schwimmer and S.~Theisen,
  ``Numerical determination of entanglement entropy for a sphere,''
  Phys.\ Lett.\  B {\bf 685}, 222 (2010)
  [arXiv:0911.4283 [hep-lat]].
}

\lref\CasiniHuertab{
  H.~Casini and M.~Huerta,
  ``A c-theorem for the entanglement entropy,''
  J.\ Phys.\ A  {\bf 40}, 7031 (2007)
  [arXiv:cond-mat/0610375].
}

\lref\CasiniHuertac{
  H.~Casini and M.~Huerta,
  ``A finite entanglement entropy and the c-theorem,''
  Phys.\ Lett.\  B {\bf 600}, 142 (2004)
  [arXiv:hep-th/0405111].
  }
\lref\FursaevSolodukhinb{
  D.~V.~Fursaev and S.~N.~Solodukhin,
  ``On one loop renormalization of black hole entropy,''
  Phys.\ Lett.\  B {\bf 365}, 51 (1996)
  [arXiv:hep-th/9412020].
}

\lref\Solodukhinb{
  S.~N.~Solodukhin,
  ``On 'Nongeometric' contribution to the entropy of black hole due to quantum
  corrections,''
  Phys.\ Rev.\  D {\bf 51} (1995) 618
  [arXiv:hep-th/9408068].
}

\lref\Solodukhinc{
  S.~N.~Solodukhin,
  ``The Conical singularity and quantum corrections to entropy of black hole,''
  Phys.\ Rev.\  D {\bf 51} (1995) 609
  [arXiv:hep-th/9407001].
}

\lref\Fursaevb{
  D.~V.~Fursaev,
  ``Black hole thermodynamics and renormalization,''
  Mod.\ Phys.\ Lett.\  A {\bf 10}, 649 (1995)
  [arXiv:hep-th/9408066].
}

\lref\Fursaevc{
  D.~V.~Fursaev,
  ``Spectral geometry and one loop divergences on manifolds with conical
  singularities,''
  Phys.\ Lett.\  B {\bf 334}, 53 (1994)
  [arXiv:hep-th/9405143].
}

\lref\CapperDuff{
  D.~M.~Capper and M.~J.~Duff,
  ``Trace anomalies in dimensional regularization,''
  Nuovo Cim.\  A {\bf 23} (1974) 173.
}

\lref\Duffrv{
  M.~J.~Duff,
  ``Twenty years of the Weyl anomaly,''
  Class.\ Quant.\ Grav.\  {\bf 11}, 1387 (1994)
  [arXiv:hep-th/9308075].
}

\lref\HenningsonSkenderisa{
  M.~Henningson and K.~Skenderis,
  ``The holographic Weyl anomaly,''
  JHEP {\bf 9807}, 023 (1998)
  [arXiv:hep-th/9806087].
}

\lref\HenningsonSkenderisb{
  M.~Henningson and K.~Skenderis,
  ``Holography and the Weyl anomaly,''
  Fortsch.\ Phys.\  {\bf 48}, 125 (2000)
  [arXiv:hep-th/9812032].
}

\lref\ArutyunovFrolov{
  G.~Arutyunov and S.~Frolov,
  ``Three-point Green function of the stress-energy tensor in the AdS/CFT
  correspondence,''
  Phys.\ Rev.\  D {\bf 60}, 026004 (1999)
  [arXiv:hep-th/9901121].

}

\lref\Klebanovetala{
  I.~R.~Klebanov, D.~Kutasov and A.~Murugan,
  ``Entanglement as a Probe of Confinement,''
  Nucl.\ Phys.\  B {\bf 796}, 274 (2008)
  [arXiv:0709.2140 [hep-th]].
}

\lref\Headricka{
  M.~Headrick,
  ``Entanglement Renyi entropies in holographic theories,''
  arXiv:1006.0047 [hep-th].
}

\lref\Takayanagiaa{S.~Ryu and T.~Takayanagi,
  ``Holographic derivation of entanglement entropy from AdS/CFT,''
  Phys.\ Rev.\ Lett.\  {\bf 96}, 181602 (2006)
  [arXiv:hep-th/0603001].
}

\lref\Takayanagiab{T.~Nishioka, S.~Ryu and T.~Takayanagi,
  ``Holographic Entanglement Entropy: An Overview,''
  J.\ Phys.\ A  {\bf 42}, 504008 (2009)
  [arXiv:0905.0932 [hep-th]].
}

\lref\Takayanagic{
  M.~Headrick and T.~Takayanagi,
  ``A holographic proof of the strong subadditivity of entanglement entropy,''
  Phys.\ Rev.\  D {\bf 76}, 106013 (2007)
  [arXiv:0704.3719 [hep-th]].
}

\lref\IyerWald{
  V.~Iyer and R.~M.~Wald,
  ``A Comparison of Noether charge and Euclidean methods for computing the
  entropy of stationary black holes,''
  Phys.\ Rev.\  D {\bf 52}, 4430 (1995)
  [arXiv:gr-qc/9503052].
}

\lref\Nelson{
  W.~Nelson,
  ``A Comment on black hole entropy in string theory,''
  Phys.\ Rev.\  D {\bf 50}, 7400 (1994)
  [arXiv:hep-th/9406011].
}

\lref\Hubenyetala{
   V.~E.~Hubeny, M.~Rangamani and T.~Takayanagi,
  ``A covariant holographic entanglement entropy proposal,''
  JHEP {\bf 0707}, 062 (2007)
  [arXiv:0705.0016 [hep-th]].
}

\lref\Holzheyetal{
  C.~Holzhey, F.~Larsen and F.~Wilczek,
  ``Geometric and renormalized entropy in conformal field theory,''
  Nucl.\ Phys.\  B {\bf 424}, 443 (1994)
  [arXiv:hep-th/9403108].
}

\lref\CalabreseCardya{
  P.~Calabrese and J.~L.~Cardy,
  ``Entanglement entropy and quantum field theory,''
  J.\ Stat.\ Mech.\  {\bf 0406}, P002 (2004)
  [arXiv:hep-th/0405152].

}

\lref\CalabreseCardyb{
  P.~Calabrese and J.~L.~Cardy,
  ``Entanglement entropy and quantum field theory: A non-technical
  introduction,''
  Int.\ J.\ Quant.\ Inf.\  {\bf 4}, 429 (2006)
  [arXiv:quant-ph/0505193].
}

\lref\Bombellietal{
  L.~Bombelli, R.~K.~Koul, J.~Lee and R.~D.~Sorkin,
  ``A Quantum Source of Entropy for Black Holes,''
  Phys.\ Rev.\  D {\bf 34}, 373 (1986).
}

\lref\Srednickia{
  M.~Srednicki,
  ``Entropy and area,''
  Phys.\ Rev.\ Lett.\  {\bf 71}, 666 (1993)
  [arXiv:hep-th/9303048].
}

\lref\Plenioetala{
  M.~B.~Plenio, J.~Eisert, J.~Dreissig and M.~Cramer,
  ``Entropy, entanglement, and area: analytical results for harmonic lattice
  systems,''
  Phys.\ Rev.\ Lett.\  {\bf 94}, 060503 (2005)
  [arXiv:quant-ph/0405142].
}

\lref\Plenioetalb{

  M.~Cramer, J.~Eisert, M.~B.~Plenio and J.~Dreissig,
  ``An entanglement-area law for general bosonic harmonic lattice systems,''
  Phys.\ Rev.\  A {\bf 73}, 012309 (2006)
  [arXiv:quant-ph/0505092].
}

\lref\Plenioetalc{
  J.~Eisert, M.~Cramer and M.~B.~Plenio,
  ``Area laws for the entanglement entropy - a review,''
  Rev.\ Mod.\ Phys.\  {\bf 82}, 277 (2010)
  [arXiv:0808.3773 [quant-ph]].
}

\lref\DasShb{
  S.~Das and S.~Shankaranarayanan,
  ``How robust is the entanglement entropy - area relation?''
  Phys.\ Rev.\  D {\bf 73}, 121701 (2006)
  [arXiv:gr-qc/0511066].
}

\lref\DasSha{M.~Ahmadi, S.~Das and S.~Shankaranarayanan,
  ``Is entanglement entropy proportional to area?''
  Can.\ J.\ Phys.\  {\bf 84}, 493 (2006)
  [arXiv:hep-th/0507228].
}

\lref\DasShc{
  S.~Das, S.~Shankaranarayanan and S.~Sur,
  ``Black hole entropy from entanglement: A review,''
  arXiv:0806.0402 [gr-qc].
}

\lref\Casinia{
  H.~Casini,
  ``Geometric entropy, area, and strong subadditivity,''
  Class.\ Quant.\ Grav.\  {\bf 21}, 2351 (2004)
  [arXiv:hep-th/0312238].
}

\lref\CHa{
  H.~Casini and M.~Huerta,
  ``Analytic results on the geometric entropy for free fields,''
  J.\ Stat.\ Mech.\  {\bf 0801}, P012 (2008)
  [arXiv:0707.1300 [hep-th]].
}

\lref\CHreview{
  H.~Casini and M.~Huerta,
  ``Entanglement entropy in free quantum field theory,''
  J.\ Phys.\ A  {\bf 42}, 504007 (2009)
  [arXiv:0905.2562 [hep-th]].
}

\lref\CHb{
  H.~Casini and M.~Huerta,
  ``Entanglement entropy for the n-sphere,''
  Phys.\ Lett.\  B {\bf 694}, 167 (2010)
  [arXiv:1007.1813 [hep-th]].

}

\lref\Wolfm{
 M.~M.~Wolf
 `` Violation of the antropic area law for Fermions,''
 Phys. Rev. Lett. {\bf 96}, 010404 (2004)
 [arXiv:quant-ph/0503219]
 }

\lref\GioevKlich{
 D.~Gioev and I.~Klich
 `` Entanglement entropy of fermions in any dimension and the Widom conjecture,''
 Phys. Rev. Lett. {\bf 96}, 100503 (2006)
 [arXiv:quant-ph/0504151]
 }

\lref\Bartheletal{
 T.~Barthel, M.~Chung and U.~Schollwoeck
 ``Entanglement scaling in critical two-dimensional fermionic and bosonic systems,''
 Phys. Rev. A {\bf 74}, 022329 (2006)
 [arXiv:cond-mat/0602077]
}

\lref\Lietal{
 W.~Li, L.~Leitian Ding, R.~Yu, T.~Roscilde and S.~Haas
 ``Scaling Behavior of Entanglement in Two- and Three- Dimensional Free Fermions,''
 Phys. Rev. B {\bf 74}, 0703103 (2007)
 [arXiv:quant-ph/0602094]
}

\lref\CasiniBW{
  H.~Casini and M.~Huerta,
  ``A finite entanglement entropy and the c-theorem,''
  Phys.\ Lett.\  B {\bf 600}, 142 (2004)
  [arXiv:hep-th/0405111].
}

\lref\CasiniES{
  H.~Casini and M.~Huerta,
  ``A c-theorem for the entanglement entropy,''
  J.\ Phys.\ A  {\bf 40}, 7031 (2007)
  [arXiv:cond-mat/0610375].
}

\lref\deBoerGX{
  J.~de Boer, M.~Kulaxizi and A.~Parnachev,
  ``Holographic Lovelock Gravities and Black Holes,''
  JHEP {\bf 1006}, 008 (2010)
  [arXiv:0912.1877 [hep-th]].
}

\lref\DehghaniManna{
  M.~H.~Dehghani and R.~B.~Mann,
  ``Lovelock-Lifshitz Black Holes,''
  JHEP {\bf 1007}, 019 (2010)
  [arXiv:1004.4397 [hep-th]].
}

\lref\DehghaniMannb{
  M.~H.~Dehghani and R.~B.~Mann,
  ``Thermodynamics of Lovelock-Lifshitz Black Branes,''
  Phys.\ Rev.\  D {\bf 82}, 064019 (2010)
  [arXiv:1006.3510 [hep-th]].
}

\lref\Chamseddine{
  A.~H.~Chamseddine,
  ``TOPOLOGICAL GAUGE THEORY OF GRAVITY IN FIVE-DIMENSIONS AND ALL ODD
  DIMENSIONS,''
  Phys.\ Lett.\  B {\bf 233}, 291 (1989).
}

\lref\Zanelli{
  J.~Zanelli,
  ``Lecture notes on Chern-Simons (super-)gravities. Second edition (February
  2008),''
  arXiv:hep-th/0502193.
}

\lref\TakayanagiL{
  T.~Azeyanagi, W.~Li and T.~Takayanagi,
  ``On String Theory Duals of Lifshitz-like Fixed Points,''
  JHEP {\bf 0906}, 084 (2009)
  [arXiv:0905.0688 [hep-th]].
}

\lref\KulaxiziParnacheva{
  M.~Kulaxizi and A.~Parnachev,
  ``Energy Flux Positivity and Unitarity in CFTs,''
  Phys.\ Rev.\ Lett.\  {\bf 106}, 011601 (2011)
  [arXiv:1007.0553 [hep-th]].
}

\lref\Solodukhinnr{
  S.~N.~Solodukhin,
  ``Entanglement Entropy in Non-Relativistic Field Theories,''
  JHEP {\bf 1004}, 101 (2010)
  [arXiv:0909.0277 [hep-th]].
}

\lref\HoyosKoroteev{
  C.~Hoyos and P.~Koroteev,
  ``On the Null Energy Condition and Causality in Lifshitz Holography,''
  Phys.\ Rev.\  D {\bf 82}, 084002 (2010)
  [Erratum-ibid.\  D {\bf 82}, 109905 (2010)]
  [arXiv:1007.1428 [hep-th]].
}

\lref\DonosGauntlett{
  A.~Donos and J.~P.~Gauntlett,
  ``Lifshitz Solutions of D=10 and D=11 supergravity,''
  JHEP {\bf 1012}, 002 (2010)
  [arXiv:1008.2062 [hep-th]].
}

\lref\Donosetal{
   A.~Donos, J.~P.~Gauntlett, N.~Kim and O.~Varela,
  ``Wrapped M5-branes, consistent truncations and AdS/CMT,''
  JHEP {\bf 1012}, 003 (2010)
  [arXiv:1009.3805 [hep-th]].
}

\lref\KachruYH{
  S.~Kachru, X.~Liu and M.~Mulligan,
  ``Gravity Duals of Lifshitz-like Fixed Points,''
  Phys.\ Rev.\  D {\bf 78}, 106005 (2008)
  [arXiv:0808.1725 [hep-th]].
}

\lref\CopseyMann{
  K.~Copsey and R.~Mann,
  ``Pathologies in Asymptotically Lifshitz Spacetimes,''
  arXiv:1011.3502 [hep-th].
}

\lref\Gregoryetala{
  R.~Gregory, S.~L.~Parameswaran, G.~Tasinato and I.~Zavala,
  ``Lifshitz solutions in supergravity and string theory,''
  arXiv:1009.3445 [hep-th].
}

\lref\Dowkera{
  J.~S.~Dowker,
  ``Hyperspherical entanglement entropy,''
  J.\ Phys.\ A  {\bf 43}, 445402 (2010)
  [arXiv:1007.3865 [hep-th]].
}

\lref\Dowkerb{
 J.~S.~Dowker,
 ``Entanglement entropy for even spheres,''
 arXiv:1009.3854 [hep-th].
 }

\lref\Dowkerc{
  J.~S.~Dowker,
  ``Determinants and conformal anomalies of GJMS operators on spheres,''
  arXiv:1010.0566 [hep-th].
}

\lref\Dowkerd{
  J.~S.~Dowker,
  ``Entanglement entropy for odd spheres,''
  arXiv:1012.1548 [hep-th].
}

\lref\MyersXS{
  R.~C.~Myers and A.~Sinha,
  ``Seeing a c-theorem with holography,''
  Phys.\ Rev.\  D {\bf 82}, 046006 (2010)
  [arXiv:1006.1263 [hep-th]].
}

\lref\MyersTJ{
  R.~C.~Myers and A.~Sinha,
  ``Holographic c-theorems in arbitrary dimensions,''
  arXiv:1011.5819 [hep-th].
}

\lref\LiuXC{
  J.~T.~Liu, W.~Sabra and Z.~Zhao,
  ``Holographic c-theorems and higher derivative gravity,''
  arXiv:1012.3382 [hep-th].
}

\lref\BuchelTT{
  A.~Buchel and R.~C.~Myers,
  ``Causality of Holographic Hydrodynamics,''
  JHEP {\bf 0908}, 016 (2009)
  [arXiv:0906.2922 [hep-th]].
}

\lref\HofmanUG{
  D.~M.~Hofman,
  ``Higher Derivative Gravity, Causality and Positivity of Energy in a UV
  complete QFT,''
  Nucl.\ Phys.\  B {\bf 823}, 174 (2009)
  [arXiv:0907.1625 [hep-th]].
}

\lref\deBoerPN{
  J.~de Boer, M.~Kulaxizi and A.~Parnachev,
  ``$AdS_7/CFT_6$, Gauss-Bonnet Gravity, and Viscosity Bound,''
  JHEP {\bf 1003}, 087 (2010)
  [arXiv:0910.5347 [hep-th]].
}

\lref\CamanhoVW{
  X.~O.~Camanho and J.~D.~Edelstein,
  ``Causality constraints in AdS/CFT from conformal collider physics and
  Gauss-Bonnet gravity,''
  JHEP {\bf 1004}, 007 (2010)
  [arXiv:0911.3160 [hep-th]].
}

\lref\BuchelSK{
  A.~Buchel, J.~Escobedo, R.~C.~Myers, M.~F.~Paulos, A.~Sinha and M.~Smolkin,
  ``Holographic GB gravity in arbitrary dimensions,''
  JHEP {\bf 1003}, 111 (2010)
  [arXiv:0911.4257 [hep-th]].
}

\lref\CamanhoHU{
  X.~O.~Camanho and J.~D.~Edelstein,
  ``Causality in AdS/CFT and Lovelock theory,''
  JHEP {\bf 1006}, 099 (2010)
  [arXiv:0912.1944 [hep-th]].
}

\lref\MyersJV{
  R.~C.~Myers, M.~F.~Paulos and A.~Sinha,
  ``Holographic studies of quasi-topological gravity,''
  JHEP {\bf 1008}, 035 (2010)
  [arXiv:1004.2055 [hep-th]].
}

\lref\MyersRU{
  R.~C.~Myers and B.~Robinson,
  ``Black Holes in Quasi-topological Gravity,''
  JHEP {\bf 1008}, 067 (2010)
  [arXiv:1003.5357 [gr-qc]].
}

\lref\KulaxiziPZ{
  M.~Kulaxizi and A.~Parnachev,
  ``Supersymmetry Constraints in Holographic Gravities,''
  Phys.\ Rev.\  D {\bf 82}, 066001 (2010)
  [arXiv:0912.4244 [hep-th]].
}

\lref\HungMyersSmolkin{
 L.~Y.~Hung, R.~C.~Myers and M.~Smolkin,
 ``On Holographic Entanglement Entropy and Higher Curvature
Gravity'',  arxiv:hep-th/1101.5813
}

\lref\CamanhoRU{
  X.~O.~Camanho, J.~D.~Edelstein and M.~F.~Paulos,
  ``Lovelock theories, holography and the fate of the viscosity bound,''
  arXiv:1010.1682 [hep-th].
}

\lref\ZamolodchikovGT{
  A.~B.~Zamolodchikov,
  ``Irreversibility of the Flux of the Renormalization Group in a 2D Field
  Theory,''
  JETP Lett.\  {\bf 43}, 730 (1986)
  [Pisma Zh.\ Eksp.\ Teor.\ Fiz.\  {\bf 43}, 565 (1986)].
}

\lref\CardyCWA{
  J.~L.~Cardy,
  ``Is There a c Theorem in Four-Dimensions?,''
  Phys.\ Lett.\  B {\bf 215}, 749 (1988).
}

\lref\JacobsonXS{
  T.~Jacobson and R.~C.~Myers,
  ``Black Hole Entropy And Higher Curvature Interactions,''
  Phys.\ Rev.\ Lett.\  {\bf 70}, 3684 (1993)
  [arXiv:hep-th/9305016].
}

\lref\BanadosQP{
  M.~Banados, C.~Teitelboim and J.~Zanelli,
  ``Black hole entropy and the dimensional continuation of the Gauss-Bonnet
  theorem,''
  Phys.\ Rev.\ Lett.\  {\bf 72}, 957 (1994)
  [arXiv:gr-qc/9309026].
}

\lref\GaiottoJF{
  D.~Gaiotto, N.~Seiberg and Y.~Tachikawa,
  ``Comments on scaling limits of 4d N=2 theories,''
  JHEP {\bf 1101}, 078 (2011)
  [arXiv:1011.4568 [hep-th]].
}

\lref\ErkalSH{
  D.~Erkal and D.~Kutasov,
  ``a-Maximization, Global Symmetries and RG Flows,''
  arXiv:1007.2176 [hep-th].
}

\lref\PolandWG{
  D.~Poland and D.~Simmons-Duffin,
  ``Bounds on 4D Conformal and Superconformal Field Theories,''
  arXiv:1009.2087 [hep-th].
}

\lref\BriganteNU{
  M.~Brigante, H.~Liu, R.~C.~Myers, S.~Shenker and S.~Yaida,
  ``Viscosity Bound Violation in Higher Derivative Gravity,''
  Phys.\ Rev.\  D {\bf 77}, 126006 (2008)
  [arXiv:0712.0805 [hep-th]].
}

\lref\BriganteGZ{
  M.~Brigante, H.~Liu, R.~C.~Myers, S.~Shenker and S.~Yaida,
  ``The Viscosity Bound and Causality Violation,''
  Phys.\ Rev.\ Lett.\  {\bf 100}, 191601 (2008)
  [arXiv:0802.3318 [hep-th]].
}

\lref\MyersPK{
  R.~C.~Myers, S.~Sachdev and A.~Singh,
  ``Holographic Quantum Critical Transport without Self-Duality,''
  arXiv:1010.0443 [hep-th].
}

\lref\BuchelWF{
  A.~Buchel and S.~Cremonini,
  ``Viscosity Bound and Causality in Superfluid Plasma,''
  JHEP {\bf 1010}, 026 (2010)
  [arXiv:1007.2963 [hep-th]].
}

\lref\PapadimitriouAS{
  I.~Papadimitriou,
  ``Holographic renormalization as a canonical transformation,''
  arXiv:1007.4592 [hep-th].
}

\lref\BalasubramanianUK{
  K.~Balasubramanian and K.~Narayan,
  ``Lifshitz spacetimes from AdS null and cosmological solutions,''
  JHEP {\bf 1008}, 014 (2010)
  [arXiv:1005.3291 [hep-th]].
}

\lref\PadmanabhanWG{
  T.~Padmanabhan,
  ``Finite entanglement entropy from the zero-point-area of spacetime,''
  Phys.\ Rev.\  D {\bf 82}, 124025 (2010)
  [arXiv:1007.5066 [gr-qc]].
}

\lref\NesterovYI{
  D.~Nesterov and S.~N.~Solodukhin,
  ``Gravitational effective action and entanglement entropy in UV modified
  theories with and without Lorentz symmetry,''
  Nucl.\ Phys.\  B {\bf 842}, 141 (2011)
  [arXiv:1007.1246 [hep-th]].
}

\lref\NesterovJH{
  D.~Nesterov and S.~N.~Solodukhin,
  ``Short-distance regularity of Green's function and UV divergences in
  entanglement entropy,''
  JHEP {\bf 1009}, 041 (2010)
  [arXiv:1008.0777 [hep-th]].
}

\lref\SunUF{
  J.~R.~F.~Sun,
  ``Note on Chern-Simons Term Correction to Holographic Entanglement Entropy,''
  JHEP {\bf 0905}, 061 (2009)
  [arXiv:0810.0967 [hep-th]].
}

\lref\BianchiKW{
  M.~Bianchi, D.~Z.~Freedman and K.~Skenderis,
  ``Holographic renormalization,''
  Nucl.\ Phys.\  B {\bf 631}, 159 (2002)
  [arXiv:hep-th/0112119].
}

\lref\BergTY{
  M.~Berg and H.~Samtleben,
  ``An Exact holographic RG flow between 2-d conformal fixed points,''
  JHEP {\bf 0205}, 006 (2002)
  [arXiv:hep-th/0112154].
}

\lref\PapadimitriouAP{
  I.~Papadimitriou and K.~Skenderis,
  ``AdS / CFT correspondence and geometry,''
  arXiv:hep-th/0404176.
}

\lref\TaylorXW{
  M.~Taylor,
  ``More on counterterms in the gravitational action and anomalies,''
  arXiv:hep-th/0002125.
}

\lref\CasiniKV{
  H.~Casini, M.~Huerta and R.~C.~Myers,
  ``Towards a derivation of holographic entanglement entropy,''
  arXiv:1102.0440 [hep-th].
}

\Title{\vbox{\baselineskip12pt
}}
{\vbox{\centerline{Holographic Entanglement Entropy in Lovelock Gravities }
\vskip.06in
}}

\centerline{Jan de Boer${}^a$, Manuela Kulaxizi${}^b$, and Andrei Parnachev${}^c$}
\bigskip
\centerline{{\it ${}^a$Institute for Theoretical Physics, University of Amsterdam,}}
\centerline{{\it Science Park 904, Postbus 94485, 1090 GL Amsterdam, The Netherlands }}
\centerline{{\it ${}^b$Theoretical Physics, Department of Physics and Astronomy, Uppsala University}}
\centerline{{\it Box 516, SE-751 20 Uppsala, Sweden}}
\centerline{{\it ${}^c$Institute of Cosmos Sciences and E.C.M., Facultat de Fisica,}}
\centerline{{\it Universitat de Barcelona, Av. Diagonal 647, 08028 Barcelona, Spain}}
\vskip.1in \vskip.1in \centerline{\bf Abstract}
\noindent We study entanglement entropies of simply connected surfaces in
field theories dual to Lovelock gravities.
We consider Gauss-Bonnet and cubic Lovelock gravities in detail.
In the conformal case the logarithmic terms in the entanglement entropy are
governed by the conformal anomalies of the CFT; we verify that the holographic calculations
are consistent with this property.
We also compute the holographic entanglement entropy of a slab in the Gauss-Bonnet examples dual to
relativistic and non-relativistic CFTs  and discuss its properties.
Finally, we discuss features of the entanglement entropy in the
backgrounds dual to renormalization group flows between fixed points and comment
on the implications for a possible c-theorem in four spacetime dimensions.

\vfill

\Date{May 2011}


\newsec{Introduction and summary}

\noindent
Entanglement entropy is an interesting non-local observable which
carries important information about field theory.
Refs. \refs{\Takayanagiaa-\Takayanagia}  proposed a way of computing
entanglement entropy in the strongly coupled conformal field theories dual to
gravitational theories whose gravity sector is described by the
Einstein-Hilbert lagrangian with the negative cosmological constant.
The set of CFTs that admit duals of this type is strongly restricted.
In particular in four spacetime dimensions, all such CFTs necessarily
have their $a$ and $c$ central charges equal to each other.

It is known that some interesting phenomena in CFTs and, more generally,
in quantum field theories, are associated with the regime where $a\neq c$.
For example, there are unitarity constraints in CFTs \HofmanMaldacena\
which restrict the ratio of $a/c$ to lie within certain bounds.
Another interesting and important question is whether one can
formulate and prove the analog of Zamolodchikov's c-theorem \ZamolodchikovGT\ in three and
more spacetime dimensions.
It has been suggested that the value of the
c-function in four dimensions is equal to $a$ at fixed points \CardyCWA;
recent work in field theory includes \refs{\ErkalSH\PolandWG-\GaiottoJF}.
The conjecture is known by the name ``a-theorem''.

Holographic theories with higher derivative terms provide a natural
arena for investigating these phenomena. Recently Myers and Sinha
\refs{\MyersXS-\MyersTJ} have shown that one can formulate an a-theorem
in certain higher derivative theories of gravity, but it is not completely clear what
the field theoretic counterpart of the corresponding a-function is.
Their results have been generalized to Lovelock theories in \LiuXC.
(Work which uses higher derivative gravitational theories to study  unitarity
constraints in CFTs includes
\refs{\BriganteNU\BriganteGZ\BuchelTT\HofmanUG\deBoerPN\CamanhoVW\BuchelSK\deBoerGX\CamanhoHU\MyersJV\KulaxiziParnacheva\BuchelWF\MyersPK-\CamanhoRU}.)
Interestingly, entanglement entropy provides an independent
way of computing the $a$ and $c$ central charges in the CFTs.
In particular, it has been noticed in \Solodukhina\ that depending on the
shape of the surface which defines entanglement entropy, the logarithmic
terms contain a linear combination of the $a$ and $c$ central charges.
This provides an additional motivation to
investigate the holographic entanglement entropy (EE) in
the theories with higher derivative gravitational terms.
In this paper we consider Lovelock  gravities, paying special attention
to the Gauss-Bonnet and cubic Lovelock cases.
We make use of the prescription of \Fursaeva\  and generalize it to the Lovelock case to compute
the logarithmic terms in the holographic entanglement entropy for a few simple geometries
such as a ball, a cylinder and a slab.

The rest of the paper is organized as follows.
In the next Section we review the results of \Solodukhina\ which
imply that EE of a ball contains a logarithmic term proportional to
the $a$ central charge, while the EE of a cylinder in the similar manner
encodes the $c$ central charge.
In Section 3 we give a brief review of Lovelock theories of gravity.
Section 4 contains the description of holographic entropy proposal
of \refs{\Takayanagiaa-\Takayanagia} together with the generalization
to the Gauss-Bonnet case \Fursaeva.
There we holographically compute the logarithmic terms in the EE of a ball and  a cylinder
in CFTs dual to Gauss-Bonnet gravity in $AdS_5$ and confirm that they are proportional to the
$a$ and $c$ central charges respectively.
We also compute the entanglement entropy of a slab as a function of Gauss-Bonnet parameter.
In Section 5 we  make an educated guess for the holographic formula valid in
all Lovelock theories.
We show that the coefficient of the logarithmic term in the EE of a cylinder in six-dimensional
CFT dual to cubic Lovelock gravity in $AdS_7$ is proportional to
the linear combination of the B-type anomaly coefficients.
In Section 6 we consider the solution in the bulk which is holographically
dual to the non-relativistic field theory with Lifshitz symmetries.
We compute the entanglement entropy of a slab and a cylinder and compare
with earlier results.
Section 7 is devoted to studying entanglement entropy in the
bulk geometries dual to renormalization group flows between conformal fixed points,
motivated by the search for a c-theorem in four spacetime dimensions.

\bigskip

\noindent {\bf Note added:} As we were working on this project, we became aware
of the forthcoming paper \HungMyersSmolkin\ which partially overlaps with our results 
(see also \CasiniKV).

\newsec{Entanglement entropy and conformal field theories in four dimensions.}

\noindent The entropy of entanglement (EE) in a $d$-dimensional
quantum field theory on $\IR_{d-1}\times \IR$ is defined as the von
Neumman entropy of the reduced density matrix associated with a
subspace $V$ of the total space $\IR_{d-1}$ where the field theory lives
\eqn\entdef{S(V)=-\tr_V{\rho_V \ln{\rho_V}} }
The EE can be ultraviolate (UV) divergent in the continuum
limit and a cutoff $\epsilon$ needs to be
introduced. The leading divergent term is usually proportional to the area
of the boundary of $V$, $(\p V)$
\eqn\eeal{S(V)\sim \gamma{Area(\p
V)\over \epsilon^{d-2}}+\OO\left({1\over\epsilon^{d-3}}\right) }
where the proportionality coefficient $\gamma$ depends on the regularization
procedure. This result, known as the `` area law" for EE, was first found
numerically \refs{\Bombellietal, \Srednickia} and later derived analytically
\refs{\Plenioetala\Plenioetalb\DasSha\DasShb\Casinia-\CHa}. Note
however that the area law is violated in the presence of a
finite Fermi surface \refs{\Wolfm\GioevKlich\Bartheletal-\Lietal}.


For $d$-dimensional conformal field theories (CFTs) the
structure of the divergent terms usually takes the following
form \CasiniHuertaa\
\eqn\ententdiv{S(V)={g_{d-2}[\p V]\over
\epsilon^{d-2}}+\cdots +{g_{1}[\p V]\over \epsilon} +g_{0}[\p
V]\ln{\epsilon}+s(V)\, .} Here, $s(V)$ is the finite part of the
entropy and $g_i[\p V]$ are local, homogeneous of degree $i$,
functions of the characteristic length scale of the boundary $(\p
V)$. Eq. \ententdiv\ is based both on the local nature of the
ultraviolate divergences and on the fact that regions with common
boundary share the same entropy. In general, the terms $g_{d-2}[\p
V],\cdots, g_1[\p V]$ are not physical and depend on the
regularization procedure. On the other hand, the coefficient of
the logarithmic term is physical and universal in nature, not
affected by cutoff redefinitions.

Here our primary focus will be on conformal field
theories in $d=4$ dimensions where the universal coefficient
of EE was recently obtained  \Solodukhina.
In particular, using the replica trick and conformal invariance
of a four dimensional CFT on a curved manifold, \Solodukhina\ proposed that
the coefficient of the logarithmically divergent term in the
entanglement entropy of a smooth and connected region $V$ is given
by
\eqn\logproposal{g_0[\p V]={c\over 720 \pi} g_{0c}[\p
V]-{a\over 720\pi} g_{0a}[\p V]} where $(c, a)$ are the central
charges of the four dimensional CFT and $g_{0c}, g_{0a}$ are defined as follows
\eqn\gzeroca{\eqalign{g_{0c}[\p V]&=\int_{\p V}
R_{\mu\nu\sigma\tau}(n_i^{\mu}n_i^{\sigma})(n_j^{\nu}n_j^{\tau})-R_{\mu\nu}n_i^{\mu}n_i^{\nu}+{1\over
3}R+ \mu\left[{1\over 2}k^i k^i-(k_{\mu\nu}^i)^2\right]\cr
g_{0a}[\p V]&=\int_{\p V}R_{(\p V)}=\int_{\p A}
R_{\mu\nu\sigma\tau}(n_i^{\mu}n_i^{\sigma})(n_j^{\nu}n_j^{\tau})-2
R_{\mu\nu}n_i^{\mu}n_i^{\nu}+R+ \left[k^i
k^i-(k_{\mu\nu}^i)^2\right]  .}}
Notice that $g_{0a}[\p V]$ is simply the Euler character of the boundary manifold $(\p V)$ which
we have also expressed (with the help of the Gauss-Codazzi identity) in terms of the ambient spacetime
Riemann $R_{\mu\nu\sigma\tau}$ and Ricci $R_{\mu\nu}$ curvatures.
$g_{0c}[\p V]$ on the other hand, to the best of our knowledge
does not have a clear geometric meaning.

Let us clarify the notation used in \gzeroca. Suppose the two dimensional boundary $(\p V)$ is parameterized
by a set of coordinates $x^a$ with $a=1,2$ whereas the spacetime metric $g_{\mu\nu}$ where the CFT lives is spanned by
coordinates $X^{\mu}$ with $\mu=0,1,2,3$. Then, $n_i^{\mu}$ in \gzeroca\ denote two $(i=1,2)$ vectors normal to the
surface $(\p V)$ satisfying:
\eqn\neqs{\eqalign{n_i^{\mu}n_j^{\nu}g_{\mu\nu}&=\delta_{ij}\cr
g_{\mu\nu}{\p X^\mu\over \p x^a} n_i^\nu&=0 }. }
$k^i_{\mu\nu}$ represents the extrinsic curvature tensor of $(\p V)$ associated to the normal $n_i$ and is given by
\eqn\extcurv{k^i_{\mu\nu}=-\gamma^\rho_\mu \gamma^\sigma_\nu \nabla_\rho n^i_{\sigma}\, ,}
where $\gamma_{\mu\nu}$ represents the induced metric equal to $\gamma_{\mu\nu}=g_{\mu\nu}-n^i_\mu n^i_\nu$.
Note that $\mu,\nu=0,1,2,3$ are spacetime indices raised and lowered with the metric $g_{\mu\nu}$.
Finally, $k^i$ is the trace of the extrinsic curvature tensor $k^i=k^i_{\mu\nu}g^{\mu\nu}$.

The coefficient $\mu$ in \gzeroca\ cannot be determined by conformal invariance. In \Solodukhina\ it is fixed by requiring agreement
between \logproposal\ and the holographic calculation.
Using this result \Solodukhina\ concluded that the coefficient of the logarithmic term in the entanglement
entropy of a ball $B$ and a cylinder $C$ in {\it{any}} four dimensional conformal field theory takes the form
\eqn\ententfoursc{\eqalign{S(B)&=\cdots+{a\over 90}\ln{\epsilon}+s(B)\cr
S(C)&=\cdots+{c\over 720}{l\over R}\ln{\epsilon}+s(C) }}
where $B, C$ denote a ball of radius $R$ and an infinite cylinder of radius $R$ and length $l$
respectively\foot{Note that for a region $V$ with zero extrinsic curvature, e.g. a slab, the logarithmic term vanishes.}.
This interesting result provides an additional characterization of the anomaly
coefficients $(c, a)$ through entanglement entropy.

Independent evidence in support of the work of \Solodukhina\ was
first given in \Lohmayera. The authors of \Lohmayera\ used
Srednicki's regularization method and numerically computed the
coefficient of the logarithmic term in the entanglement entropy of
a ball for a free bosonic CFT in $d=4$ dimensions. Their result
was in complete agreement with \Solodukhina. Recently, \CHb\
analytically computed the entanglement entropy for the region of a
ball in a massless scalar field theory in arbitrary dimensions,
further verifying Solodukhin's formula for this case. General results
for the entanglement entropy for a spherical region were established later
on in \refs{\Dowkera\Dowkerb\Dowkerc-\Dowkerd}.

It is interesting to use eqs \logproposal, \gzeroca\ to compute the
coefficients of the logarithmic term of the EE for spatial regions of
different geometrical shape such as, ellipsoids, toroids e.t.c.
It is important to stress here that the results of \Solodukhina\ are restricted
to regions $V$ of smooth geometrical shape. Otherwise, contributions from
the non-smooth boundary are likely to modify the coefficients of the
logarithmically divergent terms in the entropy \CasiniHuertaa.

\newsec{Gauss-Bonnet gravity}

\noindent Among all theories of gravity which contain higher
derivative terms of the Riemann tensor in the their action there
exists a special class of theories usually referred to, as
Lovelock gravity. This class of gravitational theories stands out
both for its simplicity and the several properties it shares with
Einstein-Hilbert gravity. In particular, it is the most general
theory of gravity whose equations of motion involve only second
order derivatives of the metric. It is ghost free when expanded
around a Minkowski spacetime background, while recently, the
Palatini and metric formulations of Lovelock gravity have been
shown to be equivalent \ExirifardJabbari.

The action for Lovelock gravity in $d+1$-dimensions is
\eqn\LovelockAction{S={1\over 16 \pi G_{N}^{d+1}} \int d^{{d+1}}x \sqrt{-g}\sum_{p=0}^{[{d\over 2}]} (-)^p {(d-2 p)!\over (d-2)!}\lambda_{p} {\cal{L}}_{p}\, ,}
where $G_{N}^{d+1}$ is the $d+1$--dimensional Newton's constant, $[{d\over 2}]$ denotes the integral part of ${d\over 2}$, $\lambda_p$
is the $p$-th order Lovelock coefficient\foot{Note that $\lambda_p$ are denoted as $\hat{\lambda}_p$ in \deBoerGX.} and ${\cal{L}}_p$ is the Euler density of a $2p$--dimensional manifold.
In $d+1$ dimensions all ${\cal{L}}_p$ terms with $p\geq [{d\over 2}]$
are either total derivatives or vanish identically.

In this work we will primarily focus on five dimensional gravitational theories and limit ourselves to the Gauss-Bonnet action.
This is the simplest example of a Lovelock action, with only the 4--dimensional Euler density included
\eqn\action{S={1\over 16 \pi G_{N}^{(5)}}\int d^{5} x  \sqrt{-g}\left(R+{12 \over L^2}+
{\lambda L^2\over 2}  {\cal{L}}_{(2)}\right)\, .}
Note that in eq. \action\ we introduced a cosmological constant term $\Lambda=-{12 \over L^2}$ and
denoted the dimensionless Gauss-Bonnet parameter by $\lambda$ instead of $\lambda_2$ since the other
Lovelock terms vanish in this case. In what follows we will retain this notation except for section 5,
where we discuss generic Lovelock theories.
The Gauss-Bonnet term ${\cal{L}}_{(2)}$ in \action\ is
\eqn\gblang{{\cal{L}}_{(2)} =R_{MNPQ} R^{MNPQ}-4 R_{MN} R^{MN}+R^2 }
Equations of motion derived from \action\  are expressed in the following way
\eqn\eqom{-{1\over 2}g_{MN}{\cal{L}}+R_{MN}+\lambda L^2{\cal{H}}^{(2)}_{MN} =0}
with ${\cal {H}}^{(2)}_{MN}$ defined as
\eqn\Hdef{{\cal{H}}^{(2)}_{MN}=R_{MLPQ} R_{N}^{\quad LPQ}-2 R_{MP}R_{N}^{\quad P}-2 R_{MPNQ}R^{PQ}+R R_{MN}}
Eq. \eqom\ admits AdS solutions of the form\refs{\BoulwareDeser-\Cai}
\eqn\bhansatz{ds^2={L_{AdS}^2 dr^2\over r^2}+{r^2\over L_{AdS}^2} \left(-dt^2+\sum_{i=1}^{d-1} dx_i dx^i\right)}
where the curvature scale of the AdS space is related to the cosmological constant via\foot{To be specific,
Gauss-Bonnet gravity admits another AdS solution with $\alpha={2\over 1-\sqrt{1-4 \lambda}}$ but this solution
is unstable and contains ghosts \BoulwareDeser.}
\eqn\alphadef{L_{AdS}={L\over\sqrt{\alpha}}\qquad\alpha={2\over 1+\sqrt{1-4 \lambda}} .}

Gauss-Bonnet gravity has been extensively studied in the context of the
AdS/CFT correspondence.
The basic aspects of the holographic dictionary established in the case of
Einstein--Hilbert gravity remain the same, since the equations of motion
retain their second order form.
Moreover, the additional parameter $\lambda$
allows for a holographic CFT dual with unequal central
charges $(c, a)$. It thus provides an opportunity to investigate several
new aspects of the correspondence (recall that all AdS backgrounds satisfying
the Einstein-Hilbert equations of motion yield $a=c$).


There are two ways to relate the gravitational parameters, the Gauss-Bonnet coupling $\lambda$,
Newton's five dimensional coupling constant $G_{N}^{(5)}$ and the cosmological constant $L$,
to the CFT parameters $(c, a)$. One is via a holographic calculation of the three point
function of the stress energy tensor and the other through the holographic computation
of the Weyl anomaly \refs{\HenningsonSkenderisa-\HenningsonSkenderisb} . Both calculations yield the same result, which is a good consistency check.
The holographic calculation of the Weyl anomaly in Gauss-Bonnet gravity was performed in \NojiriMH. Here
we simply quote the results
\eqn\cahol{\eqalign{c&=45 \pi{L_{AdS}^3\over G_N^{(5)}}\sqrt{1-4\lambda} \cr
a&=45\pi{L_{AdS}^3\over G_N^{(5)}}\left[-2+3\sqrt{1-4\lambda}\right] \, ,}}
where $L_{AdS}$ is given from \alphadef. In our conventions the CFT central charges $(c, a)$
are defined through the Weyl anomaly in the following way
\eqn\Weylb{T_{\mu}^{\mu}={1\over 64 \pi^2}{1\over 90}\left( c I-a {\cal{L}}_2\right)\, .}
It will be helpful for the calculations in the next section to have
the ratio ${L_{AdS}^3\over G_N^{(5)}}$ and the Gauss-Bonnet coefficient, $\lambda$,
expressed as functions of the central charges $(c, a)$
\eqn\Llambdaac{\eqalign{{L_{AdS}^3\over G_N^{(5)}}={1\over 90\pi} (3 c-&a) \, ,\qquad \lambda={(a-5 c)(a-c)\over 4 (a-3 c)^2}\cr
& \sqrt{1-4 \lambda}=
{2 c\over 3 c-a} \, .}}
Finally, we should note that the correspondence between the positivity of the energy flux in a CFT \HofmanMaldacena\
and causality of the boundary theory in Gauss-Bonnet gravity discussed in \refs{\BriganteNU\BriganteGZ\BuchelTT-\HofmanUG} ,
restricts the values of the Gauss-Bonnet parameter $\lambda$ to lie within the region $-{7\over 36}\leq\lambda\leq{9\over 100}$.
Similar results were obtained for generic Lovelock theories of gravity in arbitrary dimensions \refs{\deBoerPN\CamanhoVW\BuchelSK\deBoerGX\CamanhoHU-\MyersJV,\CamanhoRU}.


\newsec{Holographic Entanglement Entropy Proposal}

\noindent In the context of holography, entanglement entropy
received a lot of attention after the work of Ryu and Takayanagi
where a concrete proposal for evaluating the entanglement entropy
was set forth \refs{\Takayanagiaa,\Takayanagia}. In particular,
the authors of \refs{\Takayanagiaa,\Takayanagia} conjectured that
the entanglement entropy of a spatial\foot{Generalization to the
covariant case is discussed in \Hubenyetala . } region $V$ in a
$d$-dimensional CFT admitting a dual description in terms of
Einstein-Hilbert gravity is given by
\eqn\rtproposal{S(V)={1\over 4 G_N^{(d+1)}} \int_{\Sigma} \sqrt{\sigma} }
where $\Sigma$ is defined as the minimal area surface which asymptotes to the
boundary of the spatial region $V$, $(\p V)$. For more details the
reader is encouraged to consult \Takayanagiab.

This proposal has by now passed several tests. When for instance,
the spatial region $V$ extends to the whole of space, entanglement
entropy should coincide with statistical entropy. Indeed, at finite
temperature eq. \rtproposal\ naturally reduces to the
Bekenstein-Hawking entropy formula whereas for vanishing
temperature, the dual gravitational description contains no
horizon and the entropy vanishes as it should. Other properties of
the entanglement entropy like strong subadditivity or the fact
that $V$ and its complement $V_c$ have the same entropy, are also
satisfied by the holographic EE formula \refs{\Takayanagic}.
Moreover, precise agreement between the holographic computation and the
field theoretic one has been shown in the cases where explicit results are
available (mostly for two dimensional CFTs) \refs{\Holzheyetal\CalabreseCardya-\CalabreseCardyb}.

It is interesting to generalize \rtproposal\
to include higher derivative gravitational theories. The most natural idea
is to replace eq. \rtproposal\ with Wald's entropy formula.
In fact, for Gauss--Bonnet gravity,
such a proposal already exists in the literature \Fursaeva
\foot{For related work on holographic entanglement entropy and
higher curvature corrections see also \SunUF.}.
To be specific, the author of \Fursaeva\ suggested that the entanglement
entropy of a connected region $V$ of the dual CFT, can be computed in the case
of Gauss-Bonnet gravity through the following formula
\eqn\ententhol{S(V)={1\over 4 G_N^{(5)}} \int_{\Sigma}
\sqrt{\sigma}\left(1+\lambda L^2 R_{\Sigma} \right)\, }
Here the integral is evaluated on $\Sigma$, the three dimensional
surface which at the boundary of the holographic space reduces to
the two dimensional boundary $(\p V)$ of the region whose entropy
we want to compute and which is determined by minimizing \ententhol.
$\sigma$ in the same expression, corresponds to the determinant of
the induced metric on $\Sigma$ whereas $R_{\Sigma}$ is the induced
scalar curvature of $\Sigma$\foot{To make the variational problem well-defined
a boundary term should in principle be added in \ententhol. This term does not affect
the solution of the embedding surface but it changes the value of the action
evaluated on the solution and thus of the entanglement entropy. It turns out
however that the boundary term only modifies the leading UV-divergent term
in the entanglement entropy.}.

To summarize the main reasoning of \Fursaeva\ recall that to compute the
entanglement entropy on the CFT side,
one starts by evaluating the partition function on a $d$-dimensional
$n$-sheeted space -- formed by gluing the $n$-copies of $\IR_d$ along the
boundary $(\p V)$. This procedure produces a space $\RR_n$ with conical
singularities on the surface $(\p V)$. To evaluate the partition function
on $\RR_n$ holographically, it is necessary to identify the dual $d+1$
dimensional geometry, $\SS_n$. The latter should be a solution of the
gravitational action with non-zero cosmological constant, which asymptotes to
$\RR_n$ at the boundary. Finding such a solution is a difficult task.
Instead, \Fursaeva\ assumed that $\SS_n$ is given by a $n$-sheeted AdS$_{d+1}$
formed by gluing together $n$-copies of AdS$_{d+1}$ along a surface of
codimension two. Then, the problem essentially reduced to that of evaluating the
gravitational action functional on a space with conical singularities.
A method for performing this calculation (at least in some cases) has been
developed in \FursaevSolodukhina\ (see also
\refs{\FursaevSolodukhinb\Solodukhinb\Solodukhinc\Fursaevb\Fursaevc\NesterovJH\PadmanabhanWG-\NesterovYI}).
With the use of the above method for Einstein-Hilbert gravity, \Fursaeva\
arrived at the holographic entanglement entropy formula of Ryu and Takayanagi.
Considering Gauss-Bonnet gravity instead, leads to the modified expression
\ententhol.


Recently \Headricka\ questioned some of the assumptions that were used in \Fursaeva\
to derive \ententhol. Still, \ententhol\ remains
a reasonable generalization of \rtproposal\ to Gauss-Bonnet gravity.
First, the proposal agrees with Wald's entropy formula for AdS-Schwartzchild
black holes in Gauss-Bonnet gravity \refs{\JacobsonXS\BanadosQP\Nelson-\IyerWald}.
Therefore, whenever the spatial region $V$ coincides with the total
space, the entanglement entropy is guaranteed to be equal to the thermal
entropy. Moreover, the strong subadditivity property of EE is satisfied \Takayanagic.
Finally, the difference between \ententhol\ and \rtproposal\ is the
integral of a topological quantity \ie, the euler density in two dimensions,
just like the difference between the Einstein-Hilbert and Gauss-Bonnet
lagrangian is the Euler density term of four dimensions.

In the following we will use the proposal of \Fursaeva\ to compute the
entanglement entropy for a region bounded by ball, a cylinder and a slab.
Comparison with \ententfoursc\ will provide yet another check of
\ententhol.


\subsec{{\bf The entanglement entropy of a ball.}}

\noindent To compute the entanglement entropy of a ball of radius $R$, it is useful to parameterize the AdS space
in the following form
\eqn\adsmetrice{ds^2_{AdS}=L_{AdS}^2 \left[{d\rho^2\over 4\rho^2}+{1\over\rho}\left(-dt^2+dr^2+r^2d\Omega_2^2\right)\right]\, .}
The first step is to identify a three dimensional surface in the bulk of AdS which reduces to a sphere of radius $R$ at the boundary.
Taking into account the symmetries of the problem we see that the surface in question is determined by a single function $r(\rho)$.
With this ansatz the induced metric of the surface can be written as follows
\eqn\indmetrics{ds_{EE}^2=L_{AdS}^2 \left\{{1\over 4\rho^2}\left[1+4\rho\left({\p r\over \p\rho}\right)^2\right]d\rho^2+{r^2\over\rho} d\Omega_2^2\right\}\, .}
Using \indmetrics\ to compute the induced curvature $R_{\Sigma}$ and substituting into \ententhol\ yields
\eqn\LLs{S(B)={L_{AdS}^3 \Omega_2\over 4 G_N^{(5)} } \int d\rho {r^2\sqrt{1+4 \rho (r')^2}\over 2\rho^2} \left[1+\lambda\alpha \hat{R}\right]\, ,}
where $\alpha$ is given in \alphadef\ and $\hat{R}$, the induced scalar curvature in units of the AdS radius, is
\eqn\Rhat{\hat{R}={2\left[\rho+4\rho^2 (r')^2+4 \rho r\left(r'+8 \rho (r')^3-2 \rho r''\right)-r^2\left(3+20 \rho (r')^2+16 \rho^2 r' r''\right)\right]\over r^2\left[1+4\rho (r')^2\right]^2}.}
Eq. \LLs\ gives the equations of motion which determine $r(\rho)$.
To specify the coefficient of the logarithmic term it suffices to solve for $r(\rho)$ to the next to leading order in the neighborhood
of the boundary $\rho=0$. We find that
\eqn\rhobs{r(\rho)=R-{\rho\over 2R}+\cdots\, .}
The solution is identical in this order to the case $\lambda=0$. Substituting \rhobs\ into \LLs\ yields
\eqn\Sfinals{S(B)={L_{AdS}^3 \Omega_2\over 4 G_N^{(5)} }\int_{\epsilon^2} d\rho \left[{1- 6\lambda\alpha \over 2\rho^2}R^2-
{1-6\lambda\alpha\over 4\rho}+\OO(\rho^0)\right]\,.}
Using \Llambdaac\  and the definition of $\alpha$ from \alphadef\ we arrive at
\eqn\EEsphere{S(B)={a\over 90} { R^2\over \epsilon^2}+{a\over 90}\ln{\epsilon}+\cdots\, ,}
which is in complete agreement with \ententfoursc.

\subsec{{\bf The entanglement entropy of a cylinder.}}

\noindent Consider a three dimensional surface in AdS which reduces to a two dimensional cylindrical surface
of radius $R$ and length $l$ on the boundary of the AdS space.
It is then natural to parameterize the AdS metric as follows
\eqn\adsmetricc{ds^2_{AdS}=L_{AdS}^2 \left[{d\rho^2\over 4\rho^2}+{1\over\rho}\left(-dt^2+dz^2+dr^2+r^2d\phi^2\right)\right]\,}
The symmetries of the problem lead us to consider a surface described by a single function $r(\rho)$. The induced metric is then
\eqn\indmetricc{ds_{EE}^2=L_{AdS}^2 \left\{{1\over 4\rho^2}\left[1+4\rho\left({\p r\over \p\rho}\right)^2\right]d\rho^2+{1\over\rho} dz^2+r^2\phi^2\right\}\, .}
Plugging this ansatz into \ententhol\ yields
\eqn\LLc{S(C)={L_{AdS}^3\over 4 G_N^{(5)} } 2\pi l\int d\rho {r\sqrt{1+4 \rho (r')^2}\over 2\rho^2} \left[1+\alpha\lambda\hat{R}\right]\, ,}
where now $\hat{R}$ is the induced curvature of the surface in units of the AdS radius is
\eqn\hatRc{\hat{R}={2\left[2\rho(r'+8 \rho (r')^3-2\rho r'')-r\left(3+20 \rho (r')^2+16 \rho^2 r' r''\right)\right]\over r \left[1+4 \rho (r')^2\right]^2}\,.}
The equations of motions in the vicinity of $\rho=0$ are satisfied by\foot{Notice that the solution is again identical in this order
to the case $\lambda=0$.}
\eqn\rhobc{r(\rho)=R-{\rho\over 4R}+\cdots\,.}
Substituting \rhobc\ into \LLc\ yields
\eqn\Sfinalc{S(C)={L_{AdS}^3 \over 4 G_N^{(5)} } 2\pi l\int_{\epsilon^2} d\rho\left[{(1- 6\alpha\lambda) R \over 2\rho^2}-
{1-2\alpha\lambda\over 16 R \rho}+\OO(\rho^0)\right]\, .}
With the help of \Llambdaac\ and \hatRc\ we arrive at
\eqn\EEcylinder{S(C)={a\over 90} {2\pi R l\over 4\pi\epsilon^2}+{c\over 720}{l\over R}\ln{\epsilon}+\cdots\, .}
which agrees with \ententfoursc.

\subsec{{\bf The entanglement entropy of a slab.}}

\noindent The slab geometry corresponds to the region of space bounded by
$-{y\over 2}\leq x^1\leq{y\over 2}$ and infinitely extended along the
$x^2,\, x^3$ directions.
This is the simplest configuration to consider because of the large amount of symmetry.
Here, it is convenient to write the AdS metric as
\eqn\adsrcor{ds^2={L_{AdS}^2\over r^2}\left( \eta_{ij}dx^idx^j+dr^2\right)}
with $\eta_{ij}$ the four dimensional Minkowski metric.
The three dimensional induced surface can be parametrized by a single function $x_1(r)$ as follows
\eqn\dsslab{ds^2_{EE}={L_{AdS}^2\over r^2}\left[\left(1+x_1'(r)^2\right)dr^2+dx_2^2+dx_3^2\right]}
The induced curvature of the surface (in units of the AdS radius) is non-vanishing and equal to
\eqn\Rhatslab{\hat{R}=-2 {3+3 x_1'(r)^2+2 r x_1'(r)x_1''(r)\over \left(1+x_1'(r)^2\right)^2}}
where the primes indicate differentiation with respect to the radial coordinate.
The Lagrangian of the system is independent of $x_1(r)$ so there is a constant of motion
\eqn\eqnslab{{x_1'(r) \left(1-2\alpha\lambda+x_1'(r)^2\right)\over r^3\left(1+x_1'(r)^2\right)^{{3\over 2}}} = {1\over r_{\star}^3} }
which allows us to solve for $x_1'(r)$ exactly.
Since the theory is conformal, one can rescale the coordinate by denoting $\tau=r/r_*$.
Then  eq. \eqnslab\ can be written as
\eqn\eqnslabh{ \sqrt{h(\tau)} { \left(1-2\alpha\lambda+ h(\tau)\right)\over \left(1+h(\tau)\right)^{{3\over 2}}} = \tau^3 }
where $h(\tau)=x_1'(r)^2$.
It is easy to see that there are three solutions for $h(\tau)$ but only one of them is continuously connected with
the solution of the $\lambda=0$ case. In the following we restrict our attention to this solution\foot{We avoid
writing down the solution explicitly since it is not particularly illuminating.}.
It would be interesting to examine the other two solutions which at first glance appear to be complex valued.
We leave this analysis to future work.

We proceed to relate the constant of motion $r_{\star}$ with the width $y$ of the slab
\eqn\yhalf{{y\over 2}=r_{\star} I_0(\lambda) \qquad I_0(\lambda)=\int_0^1 d\tau \sqrt{h(\tau)}\,.}
Evaluating the action on the solution of \eqnslabh\ yields
\eqn\eeslabalm{S(\lambda)={L_{AdS}^3\over 4 G_{N}^{(5)}}  \left\{{1-6\alpha\lambda\over 2}{l^2 \over\epsilon^2}+
 \gamma(\lambda) {l^2 \over y^2}\right\} }
where
\eqn\Idef{\gamma(\lambda)= 4 I_0(\lambda)^2 \int_0^1 d\tau  \left[ {[1+h(\tau)]^{1\over2}\over\tau^3}
           \left(1{-}{2\alpha\lambda (3+3 h(\tau)+\tau h'(\tau))\over [1+h(\tau)]^2}\right)-{(1-2\alpha\lambda)\over\tau^3}   \right]  }
Recall from \Llambdaac\ that $\lambda$ can be expressed
as a function of the ratio ${a\over c}$ whereas ${L_{AdS}^3\over 4 G_{N}^{(5)}}={3 c-a\over 4\times 90\pi}$.
This allows us to
express the final result as
\eqn\EEslab{S(\lambda)={1\over 4\pi}{a\over 90} {l^2\over\epsilon^2}+\gamma\left(a,\, c\right) {l^2\over y^2}\,,}
with $\gamma\left(a,\, c\right)={3 c-a\over 4\pi\times 90} \gamma(\lambda)$.
The result of numerical integration in eq. \Idef\ is shown in Fig. 1.
\midinsert\bigskip{\vbox{{\epsfxsize=3in
        \nobreak
    \centerline{\epsfbox{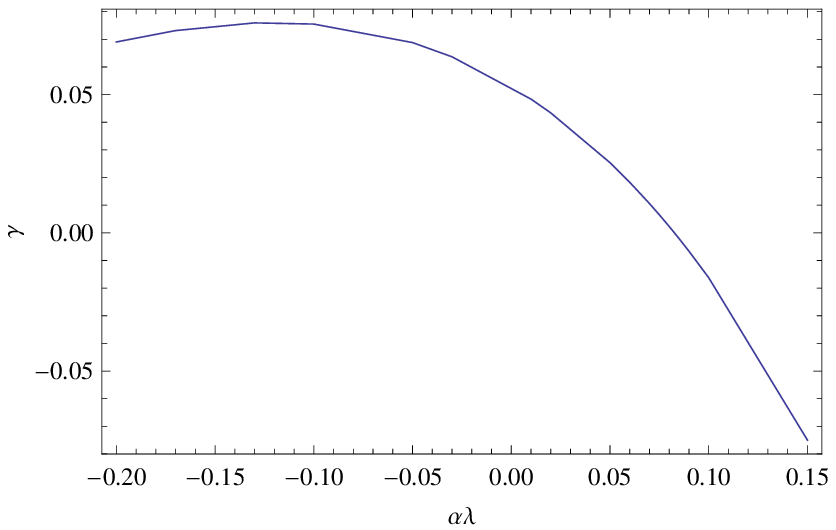}}
        \nobreak\bigskip
    {\raggedright\it \vbox{
{\bf Fig 1.}
{\it  $\gamma(\lambda)$ [see eq. \Idef]  as a function of $\lambda\alpha$.
}}}}}}
\bigskip\endinsert
\noindent
We see that the EE of a slab has the form expected from field theory considerations. The leading divergent
term follows the area law. Its coefficient is proportional to the central charge $a$ just like the entanglement entropy
of the ball and the infinite cylinder. However this coefficient depends on the regularization procedure and has no physical meaning.
On the other hand, the coefficient of the second term in \EEslab\ is universal and physical. Computing $\gamma(a,\, c)$
numerically we see that it changes sign approximately at $\alpha\lambda\sim .08$, or ${a\over c}\sim .62$. The change of sign
implies that $\gamma(a,\, c)$ is not proportional to the coefficient $\tilde{c}={(3c-a)^4\over (5 c-a)^3}$ of the
thermal entropy $s=\tilde{c}T^3$. Recall also that $\alpha\lambda={c-a\over 2(3 c-a)}$ and from Fig 1. note that
$\gamma(\lambda)$ is not linear in $(\alpha\lambda)$. This observation shows that $\gamma(a,\, c)$ is not linear
 in $a$ and $c$. It would be interesting to explore this further.
Another direction to pursue, is to consider the Gauss-Bonnet holographic dual of a confining
gauge theory. Computing the EE for a slab in this background and generalizing the results of
\Klebanovetal\ may help to clarify the meaning of $\gamma(a,\,c)$.

\newsec{Generalization to Lovelock}

\noindent So far we used Fursaev's proposal to
compute the leading divergent terms in the entanglement entropy
of a ball, an infinite cylinder and a slab in a four dimensional
CFT defined from Gauss-Bonnet gravity via gauge-gravity duality.
Gauss-Bonnet gravity allows for the holographic description of
CFTs with unequal central charges $(c, a)$ giving an opportunity
to check the proposal of \Fursaeva\ against the results of
\Solodukhina. According to \Solodukhina, the coefficient of the
logarithmic term in the entropy should be proportional to $a$ for
a ball, $c$ for an infinite cylinder and vanish for a slab,
which is precisely what we found by using the prescription of \Fursaeva.
This result led us to generalize the proposal of \Fursaeva\ to any
Lovelock theory of gravity
\eqn\eelg{S(V)={1\over 4 G_N^{(d+1)}} \sum_{p=0}^{[{d\over 2}]} (-)^{p+1} (p+1)
{(d-2 p-2)!\over (d-2)!}\lambda_{p+1}\int_{\Sigma} \sqrt{\sigma}{\cal L}_{(p)}\,.}
Expression \eelg\ coincides with the expression for the entropy of black holes
in Lovelock gravity as established in \refs{\JacobsonXS-\BanadosQP}, which
in turn agrees with Wald's entropy formula \refs{\Nelson-\IyerWald}.
As a result \eelg\ satisfies by construction several properties of the
entanglement entropy.

An interesting check of the proposal \eelg\ is the computation
of the entanglement entropy of the cylinder in the case of cubic
Lovelock gravity dual to the six-dimensional CFT.
In this case, we expect the $\log$ term in the EE to be proportional
to the coefficients of the B-type anomaly (just as in the four-dimensional
case, the coefficient was proportional to $c$) \foot{For the EE of a ball, the $\log$ term
is shown to be proportional to the coefficient of the A-type anomaly in \HungMyersSmolkin.}.
In fact, there is a non-trivial check of this statement.
In Lovelock gravity, the three coefficients of the B-type
anomaly, $b_i, i=1,\ldots,3$ are not linearly independent.
Using the results of \deBoerGX\ one can write them as
\eqn\bis{\eqalign{  b_1 & = \tc\,  {L_{AdS}^3\over4\pi G_N^{(7)} } \left( {204\over\alpha} - 168 + 100 \lambda_2 \alpha \right)\cr
                    b_2 & = \tc\,  {L_{AdS}^3\over4\pi G_N^{(7)} } \left( {75\over\alpha} - 66 + 41 \lambda_2 \alpha \right)\cr
                    b_3 & = \tc\,  {L_{AdS}^3\over4\pi G_N^{(7)} } \left( -{9\over\alpha} +6  -3  \lambda_2 \alpha \right)\cr
        }}
where we used
\eqn\elimlthree{ \lambda_3 \alpha^2=-{1\over\alpha} +1 -\lambda_2\alpha }
to eliminate terms linear in $\lambda_3$.
The overall coefficient $\tc$ is related to the definition of the invariants
$I_i, i=1,\ldots,3$ and is therefore not important.
One can fix its numerical value to be $\tc=-1/2304$.
We can now use \eelg\ to compute the EE of the cylinder.
In particular, \rhobs\ becomes
\eqn\rhobssix{  r(\rho) = R - {3\rho\over8 R}+
     {45 - 98\lambda_2\alpha -159 \lambda_2\alpha^2\over 521 (-1+2\lambda_1\alpha+3\lambda_2\alpha^2)} {\rho^2\over R^3}+\ldots }
while the result for the EE
\eqn\logcylsix{  S_{\log} =- {135\over 4 \times 512}\, {L_{AdS}^5\over 4 G_N^{(7)}} \,{2\pi^2 l\over R} \left[
              1-{94\over45} \lambda_2\alpha - {49\over 15}\lambda_3\alpha^2  \right] \ln{\epsilon}}
Using \elimlthree\ we can eliminate the term proportional to $\lambda_3$
from the square brackets.
The resulting expression has three terms, but turns out to
be proportional to the linear combination of the {\it two}
central charges \bis,
\eqn\littlemiracle{ 1-{94\over45} \lambda_2\alpha - {49\over15}\lambda_3\alpha^2 =  {19\over 360} b_1 -{1\over10} b_2   }
which is a highly nontrivial check of \eelg.
Note that the third central charge, $b_3$, is a linear combination of $b_1$ and $b_2$ (since
Lovelock gravity was shown to satisfy the supersymmetric constraint \KulaxiziPZ).
Finally, the results of this section can be helpful in extending the proposal of \Solodukhina\ to CFTs of
higher dimensionality. We hope to investigate this issue further in the future.



\newsec{Entanglement entropy in Lifshitz backgrounds.}

\noindent Lifshitz spacetimes are geometries of the form
\eqn\lifshitzmetric{ds^2={\hat{L}^2\over r^2}\left[dr^2-{1\over r^{2 w-2} } dt^2+\sum_{i=1}^{d-1}dx_i^2\right]}
which preserve the scaling symmetry
\eqn\symlif{r\rightarrow \kappa^{-1} r,\quad t\rightarrow \kappa^{w} t, \quad x^i\rightarrow \kappa x^i,\, .}
$w$ is usually referred to as the dynamical critical exponent. It is clear that any metric of the form \lifshitzmetric\
with $w\neq 1$ breaks Lorentz invariance.

Lifshitz spacetimes received considerable attention in the literature
(together with Schrodinger geometries) because they provide a natural playground
for the study of strongly coupled systems near non-relativistic critical points \refs{\KachruYH}.
Recently, Lifshitz geometries have been embedded in string theory \refs{\BalasubramanianUK\DonosGauntlett\Donosetal-\Gregoryetala}. These solutions
typically require non-trivial profile for fields other than the metric such as e.g. the dilaton.

It is an interesting fact, that Lovelock gravity admits Lifshitz solutions of arbitrary critical exponent $w$
for special values of the Lovelock parameters \refs{\DehghaniManna-\DehghaniMannb}.
This is to be contrasted with the Einstein-Hilbert case, where additional matter fields are required for the theory
to support such solutions\foot{Asymptotically Lifshitz black hole solutions however do not exist in pure Lovelock gravity \DehghaniManna.}. Restricting our attention to Gauss-Bonnet gravity, we find that \lifshitzmetric\ solves the equations of
motion \eqom\ as long as $\lambda={1\over 4}$ and $\hat{L}^2={L^2\over 2}$\foot{For completeness, we remark that
Gauss-Bonnet gravity (for particular values of the parameter $\lambda$) admits Lifshitz solutions of the most generic form
where several boundary coordinates scale like $x^i\rightarrow \kappa^w x^i$. However, the anisotropic scaling
is a relative notion and thus, in $d=4$ there are effectively only two physically distinct cases.
The first case, with a single coordinate of anisotropic scaling, is discussed above. The second case, of a Lifshitz
spacetime with two coordinates scaling anisotropically, also satisfies the equations of motion of Gauss-Bonnet gravity
as long as $\lambda={1+w+w^2\over 12 w}$ and $\hat{L}^2={1+w+w^2 \over 6}L^2$. We will not address the latter case
in the following which can be studied in a similar manner.}.
Note that when $\lambda={1\over 4}$ symmetries preserved by the gravitational action functional are enhanced to the SO$(4,2)$
group. The Lagrangian of \action\ then coincides with the Chern-Simons Lagrangian for the AdS group
and admits a natural supersymmetrization \Chamseddine.
It is perhaps worth mentioning that the AdS solution of Gauss-Bonnet gravity for $\lambda={1\over 4}$ is difficult to
interpret in the context of the AdS/CFT correspondence since the dual conformal field theory
has vanishing central charge $c$.

Here we will focus on computing the entanglement entropy holographically in $d=4$ dimensional field theories with
Lifshitz scaling. When the time coordinate is the only coordinate with a different scaling, the computation of
entanglement entropy is identical to the one in AdS spacetime. Therefore the results of the previous sections go
through unmodified as long as we make the substitution $\lambda\rightarrow{1\over 4}$ and $L_{AdS}\rightarrow {L \over \sqrt{2} }$.
In the following we will consider the more interesting case where rotational invariance is broken and instead of the
time coordinate, one of the spatial coordinates e.g. $x^1$, scales like $x^1\rightarrow \kappa^w x^1$. In particular, we will
compute the entanglement entropy of an infinitely extended slab and a cylinder.  Generalization to higher dimensions is
straightforward.

\subsec{\bf {Entanglement entropy of an infinite belt.}}

\noindent Consider double Wick rotation of \lifshitzmetric,
\eqn\lifshitzmetrica{ds^2={\hat{L}^2\over r^2}\left[dr^2+{1\over r^{2 w-2} } dx_1^2+\left(-dt^2+\sum_{i=2}^{d-1}dx_i^2\right)\right]}
Now there are two distinct orientations for the belt geometry, depending on whether its width (smallest size) extends
along the direction with $w$--scaling or not. Here we concentrate on the case
where the slab is infinitely extended along the direction with anisotropic scaling.


So let us consider a stripe, infinitely extended along
directions $x_1,x_3$ and width $y$ along $x_2$. It is convenient to parametrize the bulk surface by
a single function $x_2(r)$. The induced metric and curvature of the surface are
\eqn\indmetcurvlifa{\eqalign{ds_{EE}^2&={L^2\over 2r^2} \left[\left(1+x_2'(r)^2\right)dr^2+dx_3^2+{dx_1^2\over r^{2 w-2}}\right]\cr
\hat{R}&=-2{ \left(1+w+w^2\right)+\left(1+w+w^2\right) x_2'(r)^2+(1+w) r x_2'(r) x_2''(r)\over \left(1+x_2'(r)^2\right)^2 }
}}
Substituting \indmetcurvlifa\ into \ententhol\ results in a Lagrangian independent of $x_2(r)$ which
leads to the following equation of motion
\eqn\eqnslab{{x_2'(r) \left(1-w+x_2'(r)^2\right)\over \left(1+x_2'(r)^2\right)^{{3\over 2}}} = {r^{2+w}\over r_{\star}^{2+w}} \,.}
Note that $r_{\star}$ is related to the width of the slab through
\eqn\ydefla{{y\over 2}=r_{\star} I_0\qquad I_0=\int_0^1 d\tau \sqrt{h(\tau)}}
where the integral $I_0$ is expressed in terms of the dimensionless variable $\tau={r\over r_{\star}}$.
and the function $h(\tau)=\left({\p x_2\over \p r}\right)^2$.

Expressed in terms of $h(\tau)$ eq. \eqnslab\ has three solutions. Only one of them is real and continuously connected to the $w=1$ case.
Again we will restrict our attention to this case.
Evaluating \ententhol\ on the particular solution of \eqnslab\ yields the entanglement entropy
\eqn\eeslabla{S_{Lif.}(S)={2^{-{3\over 2}} L^3\over 4 G_N^{(5)}} \left[-w {l_1\times l_3\over\epsilon^{w+1}}-
\gamma{l_1\times l_3\over y^{w+1}}\right]}
where $\gamma$ is a numerical constant equal to $\gamma=4 I_0^2\times \left({-w^2-w\over w+1}- I\right)$.
$I_0$ is defined in \ydefla\ while $I$ denotes the following integral
\eqn\Idef{I=\int_0^1 d\tau {-w^2-w + (1-w-w^2)h(\tau)+h(\tau)^2+{1+w\over 2}\tau\dot{h}(\tau)\over \tau^3\left(1+h(\tau)\right)^{{3\over 2}} } -{-w^2-w \over \tau^3} .}

Let us discuss eq. \eeslabla. First of all, we find that the entanglement entropy is proportional to the boundary
area $l_1\times l_3$ of the belt. This is in accordance with expectations from field theory considerations \Solodukhinnr. We see that
the logarithmically divergent term is absent just like in the relativistic case. The power of the leading divergent term depends on the
critical exponent $w$ and is fixed by dimensional analysis.
The coefficient of the second term in \eeslabla\ which is physical and assumed to measure the total degrees of freedom of the system,
scales in the same way the cutoff scales. It is interesting to note that the sign of the leading divergent term is negative for $w>0$
in contrast to the relativistic case. This is not in contradiction with the results of the previous sections for Gauss-Bonnet gravity,
because for $w=1$, \ie, AdS space, the central charge $c$ of the dual theory vanishes and $a$ becomes negative (recall that Lifshitz solutions
exist for $\lambda={1\over 4}$). For this reason the case $w=1$ appears to be unphysical.
Nevertheless, the leading divergent term in the entropy does not have a physical meaning and its overall sign is immaterial. We expect
that a proper treatment of the boundary term would give a positive result but do not pursue this issue further here.
Similar results for the entanglement entropy of a slab in a Lifshitz background were obtained in \TakayanagiL.

\subsec{{\bf Entanglement entropy of an infinitely long cylinder.}}

\noindent Another interesting case to consider is that of an infinite cylinder
extended along a direction with anisotropic scaling. To make contact with sections 4.2 and 4.3
we express the Lifshitz metric in the following form
\eqn\lmetricc{ds^2= {L^2\over 2} \left[{d\rho^2\over 4\rho^2}+{1\over\rho}\left({dz^2\over \rho^{w-1}}-dt^2+dr^2+r^2 d\phi^2\right)\right]}
The induced metric and scalar curvature (in units ${L\over\sqrt{2}} =1$) of the surface are
\eqn\metricriccieelc{\eqalign{ds^2_{EE}&=\left(1+4 \rho r'(\rho)^2\right) {d\rho^2\over 4\rho^2}+{dz^2\over \rho^w}+{r(\rho)^2\over\rho} \cr
\hat{R}&={-2(1+w+w^2)r(\rho)-8\rho(2+2 w+w^2)r(\rho)r'(\rho)^2\over r(\rho)\left(1+4\rho r'(\rho)^2\right)^2}+ \cr
&+{4 \rho r'(\rho) (w+4\rho r'(\rho)^2 (1+w))-8 \rho^2 r''(\rho)(1+2(1+w)r(\rho)r'(\rho))\over r(\rho)\left(1+4\rho r'(\rho)^2\right)^2}
}}
where primes denote differentiation with respect to $\rho$.
Combining \ententhol\ and \metricriccieelc\ determines the equation of motion for $r(\rho)$ which takes
a rather complicated form but remains second order in derivatives. Studying the equation of motion in
the vicinity of the boundary $\rho=0$ we were able to eventually determine the exact solution. We find
that
\eqn\rhosol{r(\rho)=R\sqrt{1- {\rho\over R^2} } }
Note that \rhosol\ is independent of the critical exponent $w$\foot{It is interesting to expand the solution close to
the boundary $\rho=0$ as $r(\rho)=R-{\rho\over 2 R}-\cdots$. Note that the solution is not equal to next leading order
to the embedding function for the case of a cylinder in Gauss-Bonnet gravity (section 4.2 eq. \rhobc). This shows once more
that the order of limits $\lambda\rightarrow {1\over 4}$ and $w\rightarrow 1$ cannot be interchanged.}.
Evaluating the action \ententhol\ for $\lambda={1\over 4}$ on the solution \rhosol\ leads to
\eqn\eecl{S_{Lif.}(C)={2^{-{3\over 2}} L^3\over 4 G_N^{(5)}} 2\pi l_z\int_{\epsilon^2}^{\infty}d\rho {w R\over 2 \rho^{{w+3\over 2}} } \left[(w-1){\rho\over R^2}-(1+w)\right]   }
where $l_z$ regularizes the length of the cylinder.
Performing the integral in \eecl\ determines the entanglement entropy to be
\eqn\eeclfinal{S_{Lif.}(C)= {2^{-{3\over 2}} L^3\over 4 G_N^{(5)}} 2\pi l_z\left[{w R\over \rho^{w+1\over 2}} \left(1-{\rho\over R^2}\right)\right]_{\epsilon^2}^{\infty} \Rightarrow_{w>1} {2^{-{3\over 2}} L^3\over 4 G_N^{(5)}} 2\pi l_z
\left[{w R\over \epsilon^{w+1}} \left(-1+{\epsilon^2\over R^2}\right)\right] }
The computation of the entanglement entropy makes sense only for $w>1$\foot{We do not consider the case $w=1$ since its physical meaning
is ambiguous as explained earlier.}. Recall that Lifshitz geometries are unstable for $w<1$ (see for example \HoyosKoroteev).
We see a manifestation of this fact here through the computation of entanglement entropy.
As expected, the leading divergence is proportional to the area $2\pi l_z R$ whereas the scaling of the
cutoff is fixed by dimensional analysis. Note that the logarithmic divergence, characteristic
of the entanglement entropy on the cylinder in relativistic theories, is absent here. It would be interesting to examine
this point further from the field theory point of view.

\newsec{Entanglement entropy and domain wall geometries}

\noindent In an attempt to investigate aspects of the entanglement entropy
along renormalization group trajectories, we will consider here
domain wall geometries. These are asymptotically AdS spaces with
a metric of the form
\eqn\dwma{ds^2=dr^2+e^{2 A(r)}\eta_{ij}dx^{i}dx^{j}\, ,}
where $\eta_{ij}$ is the Minkowski metric of the dual quantum field theory
spacetime.
We assume here that domain wall geometries are solutions of Gauss-Bonnet
gravity with matter fields (e.g. scalars) just as they are solutions of Einstein--Hilbert
gravity with matter fields. In the spirit of the AdS/CFT correspondence they
correspond to renormalization group flow trajectories for the dual quantum field
theories. The main difference  here is that Gauss-Bonnet gravity contains an
additional dimensionless constant $\lambda$ which at the fixed points of
the flow, e.g. at $r=\infty$, is expressed in terms of
the ratio of the central charges $a,c$ of the dual CFT.

The central objective of this section is to explore how the universal and dimensionless coefficient of the logarithmic term in the entanglement
entropy changes along the renormalization group flow. The main assumption we will rely on is that the holographic
computation of entanglement entropy remains the same despite the non-trivial profile of the matter fields.
To make the comparison with the results of sections 4.1 and 4.2 straightforward, we
express the metric \dwma\ in a different coordinate system by performing the coordinate
transformation $\rho=\tilde{L}^2 e^{-{2 r\over \tilde{L}}}$ .
This way the IR region, $r=-\infty$, is mapped to $\rho=\infty$ while the
UV region, $r=\infty$, corresponds to $\rho=0$. Eq. \dwma\ is expressed in the
following form
\eqn\dwmb{ds^2=\tilde{L}^2\left[{d\rho^2\over 4 \rho^2}+{e^{2 U(\rho)}\over \rho}\eta_{ij}dx^{i}dx^{j}\right]\, ,}
with $U(\rho)$ related to $A(r)$ in \dwma\ through $U(r)\equiv A(r)-{r\over \tilde{L}}$, or
equivalently $e^{2 A(\rho)}\equiv \tilde{L}^2 {e^{2 U(\rho)}\over \rho}$. Since the metric \dwmb\ is asymptotically AdS, $e^{2 U(\rho)}$ admits an expansion near the
boundary $\rho=0$ of the form\foot{In special cases, e.g. \BergTY,
the expansion may include half integral powers of $\rho$ \refs{\TaylorXW\BianchiKW-\PapadimitriouAP}.}
\eqn\Ufg{e^{2 U(\rho)}=1+\beta_1 \rho+ \rho^2 \left(\beta_2+\beta_3\ln{\rho}+\beta_4\ln^2{\rho}\right)+\cdots\, .}
The dimensionfull coefficients $\beta_1,\beta_2,\cdots$ are determined by solving the field equations order by order in $\rho$.


\subsec{{\bf The entanglement entropy of a ball along the RG flow.}}

\noindent To compute the entanglement entropy of a ball of radius $R$ we will take steps similar
to the conformal case. Writing the domain wall metric as
\eqn\dweea{ds^2=\tilde{L}^2\left[{d\rho^2\over 4 \rho^2}+{e^{2 U(\rho)}\over \rho}\left(-dt^2+dr^2+r^2d\Omega_2^2\right)\right]\, ,}
we see that the ansatz $r(\rho)$ is still natural. The induced metric is then
\eqn\dwainduced{ds_{EE}^2=\tilde{L}^2 \left\{ {1\over 4\rho^2}
\left[1+4\rho e^{2 U(\rho)}\left({\p r\over \p\rho}\right)^2\right]d\rho^2+{r^2(\rho)e^{2 U(\rho)}\over\rho} d\Omega_2^2\right\} ,}
and substituting into \ententhol\ we arrive at
\eqn\LLdws{S(B)={\tilde{L}^3 \Omega_2\over 4 G_N^{(5)} } \int d\rho {r^2(\rho) e^{2 U(\rho)}\sqrt{1+4e^{2 U(\rho)} \rho (r')^2}\over 2\rho^2} \left[1+\alpha\lambda \hat{R}\right]}
where $\hat{R}=\hat{R}(r(\rho),U(\rho))$ is the scalar curvature of the induced three dimensional surface in units
of the asymptotic AdS radius in the UV, $\tilde{L}$.
The equation of motion for $r(\rho)$ derived from \LLdws\ is rather complicated. In the vicinity of the boundary
however it is solved by $r(\rho)=R-{\rho\over 2R}+\cdots$ exactly like the case $\lambda=0,\,U(\rho)=0$.
Knowledge of the near boundary behavior of $r(\rho)$ together with \Ufg\ allows us to
determine the divergent terms in the entanglement entropy of a ball
\eqn\Sfinaldws{\eqalign{S(B)&={\tilde{L}^3 \Omega_2\over 4 G_N^{(5)} }\int_{\epsilon^2} d\rho \left[{1- 6\alpha\lambda \over 2\rho^2}R^2-
{1-6\alpha\lambda\over 4\rho} \left(1-2 \beta_1 R^2 {1-2\alpha\lambda\over 1-6\alpha\lambda}\right)+\cdots\right]=\cr
\quad &={\tilde{L}^3 \left(1-6\alpha\lambda\right) \over 4 G_N^{(5)}} {R^2\Omega_2\over 2\epsilon^2} +{\tilde{L}^3\Omega_2 \over 4 G_N^{(5)}} {1-6\alpha\lambda\over 2} \left(1-2 \beta_1 R^2 {1-2\alpha\lambda\over 1-6\alpha\lambda}\right)\ln{\epsilon}+\cdots \, .  }}
Here $\alpha={L^2\over\tilde{L}^2}$. To leading order in the vicinity of $\rho=0$ it is given by \alphadef.

The coefficient of the logarithmically divergent term is modified compared to the pure AdS case by the overall factor
$\left(1-2 \beta_1 R^2 {1-2\alpha\lambda\over 1-6\alpha\lambda }\right)$ and $\alpha$ given in \alphadef.
Since ${1-2\alpha\lambda\over 1-6\alpha\lambda}$ is positive, the behavior of the coefficient of the logarithmic
term under rescalings of the radius of the ball $R\rightarrow \Lambda R$ depends on the sign of $\beta_1$.


\subsec{{\bf The entanglement entropy of a cylinder along the RG flow.}}

\noindent The case of a cylindrical surface is dealt with in a similar manner.
We write the domain wall solution as
\eqn\dweeb{ds^2={\tilde{L}}^2\left[{d\rho^2\over 4 \rho^2}+{e^{2 U(\rho)}\over \rho}\left(-dt^2+dz^2+dr^2+r^2d\phi^2\right)\right]\, ,}
and select an ansatz of the form $r(\rho)$ which leads to the following induced metric
\eqn\dwbinduced{ds_{EE}^2={\tilde{L}}^2 \left\{ {1\over 4\rho^2}
\left[1+4\rho e^{2 U(\rho)}\left({\p r\over \p\rho}\right)^2\right]d\rho^2+{e^{2 U(\rho)}\over\rho} \left(dz^2+r^2(\rho)d\phi^2\right)\right\} .}
We subsequently substitute into \ententhol\ and derive an equation of motion for $r(\rho)$ which
we solve near the boundary $\rho=0$. The solution $r(\rho)=R-{\rho\over 4R}+\OO(\rho^2)$, together with \Ufg, help
us obtain the divergent terms for the entanglement entropy of a cylinder
\eqn\Sfinaldws{\eqalign{S(C)&={{\tilde{L}}^3 \over 4 G_N^{(5)} } 2\pi l\int_{\epsilon^2} d\rho \left[{\left(1- 6\alpha\lambda\right) R\over 2\rho^2}-{1-2\alpha\lambda\over 16\rho R} \left(1-8 \beta_1 R^2\right)+\cdots\right]=\cr
\quad &={L_{AdS}^3 \left(1-6\alpha\lambda\right) \over 4 G_N^{(5)}} {2\pi l R\over 2\epsilon^2} +{L_{AdS}^3\over 4 G_N^{(5)}} {2\pi l\over R}{1-2\alpha\lambda\over 8} \left(1-8 \beta_1 R^2\right) \ln{\epsilon}+\cdots \, .  }}

Notice that the coefficient of the logarithmic term differs by a factor equal to $\left(1-8 \beta_1 R^2\right)$ compared to the conformal case.
Depending on the sign of $\beta_1$, the overall coefficient behaves in exactly the same way for
both the ball and the cylinder. It will be interesting to understand the implications of this statement.

\subsec{\bf $\beta_1$ and the weak energy condition. }

\noindent Here we will examine the weak (or null) energy condition and investigate whether it is
possible to constrain the sign of $\beta_1$ without further specifying the geometry \dwmb.
The null--energy condition implies that the matter stress energy tensor $T_{MN}$ satisfies $T_{MN}\zeta^{M}\zeta^N\geq 0$
for any arbitrary null vector $\zeta^{M}$.
Consider the most general null vector $\zeta^{M}$
\eqn\nullz{g_{\mu\nu}\zeta{^\mu}\zeta^{\nu}=0 \Rightarrow
\left(\zeta^{\rho}\right)^2=-4\rho e^{2U(\rho)}\left(\eta_{ij}\zeta^{i}\zeta^{j}\right) \, ,}
where $\eta_{ij}$ represents the four dimensional Minkowski metric.
It directly follows from \nullz\ that  $\left(\eta_{ij}\zeta^{i}\zeta^{j}\right)\leq 0$.
Substituting eq. \nullz\ into the null energy condition yields
\eqn\nullT{T_{\mu\nu}\zeta^{\mu}\zeta^\nu\geq 0 \Rightarrow -\left(\eta_{ij}\zeta^i\zeta^j\right) \left(T_{tt}+4\rho e^{2 U(\rho)}T_{\rho\rho}\right)\geq 0\, .}
Here\foot{Note that \nullT\ can also be written as $-{e^{2 U(\rho)}\over\rho} \left(\eta_{ij}\zeta^i\zeta^j\right) \left(-T_t^t+T_\rho^\rho\right)$. } we used $T_{x^i x^i}=-T_{tt}$ which follows from the symmetries of the metric \dwmb.
As we saw earlier $-\eta_{ij}\zeta^i\zeta^j \geq 0$ which results in $\left(T_{tt}+4\rho e^{2 U(\rho)}T_{\rho\rho}\right)\geq 0$.

Recall that the equations of motion \eqom\ for $d=4$ in the presence of matter reduce to
\eqn\eqomm{-{1\over 2}g_{MN}{\cal{L}}+R_{MN}+\lambda L^2{\cal{H}}^{(2)}_{MN}=\left(8\pi G_{N}^{(5)}\right) T_{MN} \, .}
Eq. \eqomm\ allows the determination of $\left(T_{tt}+4\rho e^{2 U(\rho)}T_{\rho\rho}\right)$ for arbitrary matter sector
\eqn\nullTs{\eqalign{T_{tt}+4\rho e^{2 U(\rho)}T_{\rho\rho}=&
-12 e^{2 U(\rho)} \left(U'(\rho)+\rho U''(\rho)\right)\times\cr
&\times\left(1-2\lambda\alpha+8 \lambda\alpha \rho U'(\rho)\left[1-\rho U'(\rho)\right]\right) }\, .}
The energy condition implies that the left hand side is positive for any $\rho$, therefore for $\rho=0$ as well.
The Fefferman--Graham expansion \Ufg\ helps us evaluate \nullTs\ at $\rho=0$ and combined with the null--energy condition leads to
\eqn\nullTf{6 \beta_1{\alpha-2  \over\alpha}\geq 0 \, ,}
with $\alpha=\left({L\over \tilde{L}}\right)^2$. To leading order close to the boundary $\alpha$ is defined by \alphadef.
For the stable AdS solution of Gauss-Bonnet gravity $(\alpha-2)>0$ fixes the sign of $\beta_1$ to be negative.
This is exactly what happens in Einstein-Hilbert gravity with a cosmological constant. It is interesting that had we
chosen the unstable AdS solution we would have a positive $\beta_1$.

We conclude that the coefficient of the logarithmically divergent term in the entropy of both the ball and the infinite cylinder
is monotonically increasing with dilatations (as long as $\beta_1$ is non-vanishing). Unfortunately, this does not mean much.
To apply the reasoning of \refs{\CasiniBW-\CasiniES} the full result, not just the logarithmic term, is required;
mainly because away from the conformal fixed point the EE depends on two, rather than one length scale, \ie, the radius $R$
of the ball and the scale of the theory $\mu$.



\subsec{\bf Entanglement entropy of the ball and the c-theorem in Einstein-Hilbert gravity.}

In the previous section we saw that the coefficient of the logarithmically divergent term
in the entanglement entropy depends only on the UV data of the theory. It is therefore clear
that this coefficient is not a good candidate for a function decreasing along the RG flow and
being equal to $a$ at the fixed points. This result, together with the work of Casini and Huerta
\CasiniHuertaa\ in two-dimensional field theories, lead us to consider instead the following quantity
\eqn\cat{Q\equiv -R {\p S_{EE, reg}\over \p R} \rightarrow a_{CFT} .}
Here $S_{EE, reg}$ contains only the finite part of the entanglement entropy of a
ball of radius $R$.
We would like to examine the monotonicity properties of $Q(R)$ holographically. A priori, this does
not seem an easy task since we cannot compute \cat\ exactly in an arbitrary domain wall geometry.
However, it is possible that the null energy condition and general characteristics of the spacetime \dwmb\
will determine whether \cat\ behaves monotonically along renormalization group trajectories.

In the following we will attack this problem in the context of Einstein-Hilbert
gravity. This has the advantage of being technically simpler while the relevant
features of the problem remain the same. To simplify the analysis we will parametrize
the surface in the bulk by $\rho(x^i)$ where $x^i$ for $i=1,2,3$ are cartesian coordinates at
the boundary. This choice yields a lagrangian independent of $x^i$
\eqn\actionf{\eqalign{S(B)&={\tilde{L}^3\over 4 G_N^{(5)}}\int\prod_{i=1}^3 dx_i \LL\left(\rho(x^i),\p_i\rho(x^i)\right)\cr
\LL&={e^{2 U(\rho)}\over 2 \rho(x^i)^2}\sqrt{\sum_i\left[\p_i\rho(x^i)\right]^2+4 e^{2 U(\rho)} \rho(x^i) } }}
where $\rho(x^i)$ determines the profile of the embedding surface.
Translational invariance implies that the system has a conserved "stress-energy" tensor $T_i^j$
\eqn\tijdef{T_i^j={\p \LL\over\p (\p_j\rho)} (\p_i\rho) -\LL \delta_i^j}
obeying $\p_i T_i^j=0$.
Due to spherical symmetry $T_{ij}$ takes the form
\eqn\tij{T^{ij}= A(r)\hat{r}^i\hat{r}^j+\delta^{ij} B(r)}
where $A, \, B$ can be explicitly found to be
\eqn\alphabeta{\eqalign{A(r)&={ e^{2 U(\rho)}\over 2 \rho^2} {\rho'(r)^2\over\sqrt{\rho'(r)^2+4 e^{2 U(\rho)} \rho}}=
{ e^{2 U(\rho)}\over 2 \rho^2} {1\over |r'(\rho)|\sqrt{1+4 e^{2 U(\rho)} \rho r'(\rho)^2}} \cr
B(r)&=-{e^{2 U(\rho)}\over 2 \rho^2}\sqrt{\rho'(r)^2+4 e^{2 U(\rho)} \rho r'(\rho)^2}=
-{e^{2 U(\rho)}\over 2 \rho^2|r'(\rho)|}\sqrt{1+4 e^{2 U(\rho)} \rho r'(\rho)^2 } }}
In the first equality we have replaced the dependence of $\rho$ on the cartesian coordinates $x^i$ with the spherical coordinate $r$.
In the second equality we give $A(r)\, ,B(r)$ in terms of $r(\rho)$ instead of $\rho(r)$.
Expressing the geometry of the embedding surface with $r(\rho)$ will be more convenient in the following.
Note that $r(\rho)$ satisfies the conservation equation of $T^{ij}$, in other words the equations of motion
derived from \actionf. The conservation equation can be written in a simple form in terms of the functions
$A(r)$ and $B(r)$
\eqn\conservationeqn{2 A(r)+r {d \,\over d r } \left(A(r)+B(r)\right)=0}
Solving for $r(\rho)$ in the vicinity of the UV with the help of \Ufg\ and \alphabeta\ yields
\eqn\ruv{r(\rho)=R-{\rho\over 2 R}+ \sigma_1 \rho^2+{\beta_1\over 4 R} \rho^2\log{\rho}+ \sigma_3 \rho^3 \cdots \,.}
where $\sigma_1$ is a function of the parameters $(\beta_1,\beta_2,\beta_3)$ which specify the domain wall geometry
but also of $\sigma_3$, the coefficient of a higher order term in the near boundary expansion of $r(\rho)$
\eqn\done{\sigma_1={9-27 R^2 \beta_1+4 R^4 (2 \beta_1^2+9 \beta_2-6 \beta_3)-72 R^5 \sigma_3\over 12 R^3 (-9+8 R^2 \beta_1)}\, .}
Unfortunately, $\sigma_1$ cannot be determined without knowledge of the exact form of
the profile function $r(\rho)$. This will turn out to be the main difficulty in determining
the behavior of $Q(R)$ under rescalings of $R$.

Now suppose we vary the boundary conditions in the entanglement entropy computation as follows
\eqn\varybc{\rho(x^i+\delta x^i)=\epsilon+\delta\epsilon}
where the original boundary condition is $\rho(x^i)=\epsilon$ for $x^i\in \DD$ (here $\DD$ is a sphere of radius $R$).
The change in the on-shell action would then be equal to
\eqn\varyaction{\delta S_{on-shell}={\tilde{L}^3\over 4 G_N^{(5)}}\left[
\int_{\DD} d\Sigma_i {\delta \LL\over \delta \p_i\rho}\delta\epsilon -\int_{\DD} d\Sigma_j T_i^j \delta x^i\right]}
with $d\Sigma_i$ the volume element on $\DD$. Substituting \tij\ into \varyaction\ leads to
\eqn\Qeq{Q=-R {\p S_{EE, reg}\over \p R}= ={\tilde{L}^3\Omega_2\over 4 G_N^{(5)}} R^3\left(A(R)+B(R)\right)|_{reg.} }
Having obtained the explicit form of the divergences in the previous section, it is easy to
subtract them from $A(R)+B(R)$ to arrive at
\eqn\Qres{Q=-R {\p S_{EE, reg}\over \p R}={\tilde{L}^3\Omega_2\over 4 G_N^{(5)}}{1\over 2}\left(1-3 R^2 \beta_1+8 R^3 \sigma_1\right)}
Although we have obtained a simple expression for $Q(R)$, we cannot
determine whether it has a monotonic behavior under rescalings of the radius $R$.
It appears that one would need to know the exact solution for the profile $r(\rho)$.
Smoothness and other generic characteristics of the geometry of the induced surface did
not suffice to prove the monotonicity of $Q(R)$.

Moreover, it is not clear how the null energy condition, pertinent to the ambient spacetime,
will affect the behavior of $Q(r)$ which depends on the details of the embedding function $r(\rho)$. It is
likely that even if $Q(R)$ is a c-function in the sense of Zamolodchikov,
it is most likely a different c-function than the standard holographic one -- which in the coordinates used
here is expressed as $c_{hol}\sim {1\over \left( -2 U'(\rho)\rho+1\right)^3}$.

Finally, the two main ingredients in the proof of \refs{\CasiniBW -\CasiniES} in two dimensions, are Lorentz
invariance and the strong subadditivity property of the entanglement entropy. To make use of the latter, it might
be more appropriate to consider the geometry of an annulus. We leave this issue to future investigation.

\bigskip
\bigskip

\noindent {\bf Acknowledgements:}
We would like to thank R. Myers, I. Papadimitriou, M. Rangamani  and M. Taylor
for useful discussions and H. Casini, M. Huerta and particularly S. Solodukhin
for helpful correspondence.
We are grateful to the organizers of the ``Crete Conference on Gauge
Theories and Structure of Spacetime'' for a great working atmosphere.
M.K. thanks Galileo Galilei Institute in Florence, and University of Barcelona
and A.P. thanks Galileo Galilei Institute in Florence, University of Chicago
and Uppsala University for hospitality during the completion of this work.
The work of M.K. is supported by the G{\"{o}}ran Gustaffson Foundation.



\footatend\vfill\supereject\immediate\closeout\rfile\writestoppt
\baselineskip=14pt\centerline{{\bf References}}\bigskip{\frenchspacing%
\parindent=20pt\escapechar=` \input refs.tmp\vfill\eject}\nonfrenchspacing
\end